\documentclass[aps,prd,longbibliography,reprint,amsmath,amssymb,amsfonts]{revtex4-1}
\pdfoutput=1
\usepackage[colorlinks,linkcolor=blue,urlcolor=blue,citecolor=blue,bookmarks,bookmarksnumbered]{hyperref}
\usepackage{natbib}
\usepackage{bm}
\usepackage{dcolumn}
\usepackage{graphicx}
\usepackage{setspace}
\usepackage{subfig}
\usepackage[utf8]{inputenc}

\begin{document}

\title{Jet Tagging via Particle Clouds}

\author{Huilin Qu}
\email{hqu@ucsb.edu}
\affiliation{Department of Physics, University of California, Santa Barbara, California 93106, USA}
\author{Loukas Gouskos}
\email{loukas.gouskos@cern.ch}
\affiliation{CERN, CH-1211 Geneva 23, Switzerland}

\begin{abstract}

How to represent a jet is at the core of machine learning on jet physics. Inspired by the notion of point clouds, we propose a new approach that considers a jet as an unordered set of its constituent particles, effectively a ``particle cloud''. Such a particle cloud representation of jets is efficient in incorporating raw information of jets and also explicitly respects the permutation symmetry. Based on the particle cloud representation, we propose ParticleNet, a customized neural network architecture using Dynamic Graph Convolutional Neural Network for jet tagging problems. The ParticleNet architecture achieves state-of-the-art performance on two representative jet tagging benchmarks and is improved significantly over existing methods.

\end{abstract}

\maketitle

\section{Introduction}
\label{sec:introduction}

A jet is one of the most ubiquitous objects in proton-proton collision events at the LHC. In essence, a jet is a collimated spray of particles. It serves as a handle to probe the underlying elementary particle produced in the hard scattering process that initiates the cascade of particles contained in the jet. 

One of the most important questions about a jet is which type of elementary particle initiates it. Jets initiated by different particles exhibit different characteristics. For example, jets initiated by gluons tend to have a broader energy spread than jets initiated by quarks. High-momentum heavy particles (e.g., top quarks and W, Z, and Higgs bosons) that decay hadronically can lead to jets with distinct multi-prong structures. Therefore, the identity of the source particle can be inferred from properties of the reconstructed jet. Such particle identity information provides powerful insights into the collision events under study and therefore can help greatly in separating events originating from different physics processes and improving the sensitivity of both searches for new particles and measurements of the standard model processes.

The study on jet tagging, i.e., the identification of the elementary particle initiating a jet, has a long history. Methods based on the QCD theory have been proposed and continuously improved for discriminating quark and gluon jets \cite{Gallicchio:2011xq,Gallicchio:2012ez,Larkoski:2014pca,Bhattacherjee:2015psa,FerreiradeLima:2016gcz,Gras:2017jty,Frye:2017yrw}, tagging jets originating from high-momentum heavy particles \cite{Kaplan:2008ie,Cui:2010km,Plehn:2011sj,Soper:2012pb,Anders:2013oga,Kasieczka:2015jma,Thaler:2010tr,Thaler:2011gf,Larkoski:2013eya,Moult:2016cvt,Larkoski:2014wba}, etc. See Refs. \cite{Abdesselam:2010pt,Altheimer:2012mn,Altheimer:2013yza,Adams:2015hiv,Larkoski:2017jix,Asquith:2018igt} for more in-depth reviews. Recently, machine learning (ML) has injected fresh blood in jet tagging. Jets are regarded as images \cite{Cogan:2014oua,Almeida:2015jua,deOliveira:2015xxd,Baldi:2016fql,Barnard:2016qma,Komiske:2016rsd,ATL-PHYS-PUB-2017-017,Kasieczka:2017nvn,Macaluso:2018tck,Choi:2018dag,Dreyer:2018nbf} or as sequences \cite{Guest:2016iqz,Pearkes:2017hku,Egan:2017ojy,Fraser:2018ieu,CMS-DP-2017-013,CMS-DP-2018-058,CMS-DP-2017-049,stoye2017deepjet,CMS-PAS-JME-18-002,ATL-PHYS-PUB-2017-003,Butter:2017cot,Kasieczka:2018lwf,Erdmann:2018shi}, trees \cite{Louppe:2017ipp,Cheng:2017rdo}, graphs \cite{henrionneural}, or sets \cite{Komiske:2018cqr} of particles, and ML techniques, most notably deep neural networks (DNNs), are used to build new jet tagging algorithms automatically from (labeled) simulated samples or even (unlabeled) real data \cite{Metodiev:2017vrx,Komiske:2018oaa,Andreassen:2018apy,Komiske:2018vkc}, leading to new insights and improvements in jet tagging. 

In this paper, we propose a new deep-learning approach for jet tagging using a novel way to represent jets. Instead of organizing a jet's constituent particles into an ordered structure (e.g., a sequence or a tree), we treat a jet as an \textit{unordered} set of particles \footnote{The idea of regarding jets as unordered sets of particles was also proposed in Ref. \cite{Komiske:2018cqr} independently while this work was being finalized. We provide comparison to their approach in later sections.}. This is very analogous to the point cloud representation of three-dimensional (3D) shapes used in computer vision, where each shape is represented by a set of points in space, and the points themselves are also unordered. Therefore, a jet can be viewed as a ``particle cloud''. Based on Dynamic Graph Convolutional Neural Network (DGCNN) \cite{DBLP:journals/corr/abs-1801-07829}, we design ParticleNet, a customized neural network architecture that operates directly on particle clouds for jet tagging. The ParticleNet architecture is evaluated on two jet tagging benchmarks and is found to achieve significant improvements over all existing methods.

\section{Jet representations}
\label{sec:representations}

The efficiency and effectiveness of ML techniques on jet physics relies heavily on how a jet is represented. In this section, we review the mainstream jet representations and introduce the particle cloud representation.

\subsection{Image-based representation}

The image representation has its root in the reconstruction of jets with calorimeters. A calorimeter measures the energy deposition of a jet on fine-grained spatial cells. Treating the energy deposition on each cell as the pixel intensity naturally creates an image for a jet. When jets are formed by particles reconstructed with the full detector information (e.g., using a particle-flow algorithm \cite{Sirunyan:2017ulk,Aaboud:2017aca}), a jet image can be constructed by mapping each particle onto the corresponding calorimeter cell, and sum up the energy if more than one particle is mapped to the same cell. 

The image-based approach has been extensively studied for various jet tagging tasks, e.g., $W$ boson tagging \cite{Cogan:2014oua,Almeida:2015jua,deOliveira:2015xxd,Baldi:2016fql,Barnard:2016qma,Dreyer:2018nbf}, top tagging \cite{Kasieczka:2017nvn,Macaluso:2018tck,Choi:2018dag} and quark-gluon tagging \cite{Komiske:2016rsd,ATL-PHYS-PUB-2017-017}. Convolutional neural networks (CNNs) with various architectures were explored in these studies, and they were found to achieve sizable improvement in performance compared to traditional multivariate methods using observables motivated by QCD theory. However, the architectures investigated in these papers are in general much shallower compared to state-of-the-art CNN architectures used in image classification tasks (e.g., ResNet \cite{he2016deep} or Inception \cite{szegedy2016rethinking}); therefore, it remains to be seen that if deeper architectures can further improve the performance. 

Despite the promising performance, the image-based representation has two main shortcomings. While it can include all information without loss when a jet is measured by only the calorimeter, once the jet constituent particles are reconstructed, how to incorporate additional information of the particles is unclear, as it involves combining non-additive quantities (e.g., the particle type) of multiple particles entering the same cell. Moreover, treating jets as images also leads to a very sparse representation: a typical jet has $\mathcal{O}(10)$ to $\mathcal{O}(100)$ particles, while a jet image typically needs $\mathcal{O}(1000)$ pixels (e.g., $32\times32$) in order to fully contain the jet; therefore, more than 90\% of the pixels are blank. This makes the CNNs highly computationally inefficient on jet images.

\subsection{Particle-based representation}

A more natural way to represent a jet, when particles are reconstructed, is to simply view the jet as a collection of its constituent particles. This approach allows for the inclusion of any kind of features for each particle and therefore is significantly more flexible than the image representation. It is also much more compact compared to the image representation, though at the cost of being variable length, as each jet may contain a different number of particles. 

A collection of particles, though, is a rather general concept. Before applying any deep-learning algorithm, a concrete data structure has to be chosen. The prevailing choice is a sequence, in which particles are sorted in a specific way (e.g., with decreasing transverse momentum) and organized into a one-dimensional (1D) list. Using particle sequences as inputs, jet tagging tasks have been tackled with recurrent neural networks (RNNs) \cite{Guest:2016iqz,ATL-PHYS-PUB-2017-003,Pearkes:2017hku,Egan:2017ojy,Fraser:2018ieu}, 1D CNNs \cite{CMS-DP-2017-013,CMS-DP-2018-058,CMS-DP-2017-049,stoye2017deepjet,CMS-PAS-JME-18-002} and physics-oriented neural networks \cite{Butter:2017cot,Kasieczka:2018lwf,Erdmann:2018shi}. Another interesting choice is a binary tree, which is well motivated from the QCD theory perspective. Recursive neural networks (RecNNs) are then a natural fit and have been studied in Refs. \cite{Louppe:2017ipp,Cheng:2017rdo}. 

One thing to note about the sequence or tree representation is that they both need the particles to be sorted in some way, as the order of the particles is used implicitly in the corresponding RNNs, 1D CNNs, or the RecNNs. However, the constituent particles in a jet have no intrinsic order; thus, the manually imposed order may turn out to be suboptimal and impair the performance.

\subsection{Jet as a particle cloud}

An even more natural representation than particle sequences or trees would be an unordered, permutation-invariant \textit{set} of particles. As a special case of the particle-based representations, it shares all the advantages of particle-based representations, especially the flexibility to include arbitrary features for each particle. We refer to such representation of a jet as a \textit{particle cloud}, analogous to the point cloud representation of 3D shapes used in computer vision. They are actually highly similar, as both are essentially unordered sets of entities distributed irregularly in space. In both clouds, the elements are not unrelated individuals but are rather correlated, as they represent higher-level objects (i.e., jets or 3D shapes) that have rich internal structures. Therefore, deep-learning algorithms developed for point clouds are likely to be helpful for particle clouds, i.e., jets, as well.

The idea of regarding jets as unordered sets of particles was also proposed in Ref. \cite{Komiske:2018cqr} and is in parallel to our work. The Deep Sets framework \cite{zaheer2017deep} was adapted to construct the infrared and collinear safe Energy Flow Network and the more general Particle Flow Network. However, different from the DGCNN \cite{DBLP:journals/corr/abs-1801-07829} approach adopted in this paper, the Deep Sets approach does not explicitly exploit the local spatial structure of particle clouds, but only processes the particle clouds in a global way.
Another closely related approach is to represent a jet as a graph whose vertices are the particles. Message-passing neural networks (MPNNs) with different variants of adjacency matrices were explored on such jet graphs and were found to show better performance than the RecNNs \cite{henrionneural}. However, depending on how the adjacency matrix is defined, the MPNNs may not respect the permutation symmetry of the particles.

 \section{Network architecture}
\label{sec:model}

The permutation symmetry of the particle cloud makes it a natural and promising representation of jets. However, to achieve the best possible performance, the architecture of the neural network has to be carefully designed to fully exploit the potential of this representation. In this section, we introduce ParticleNet, a CNN-like deep neural network for jet tagging with particle cloud data. 

\subsection{Edge convolution}

\newcommand\BigBox{\vcenter{\hbox{\scalebox{2}{$\Box$}}}}
\newcommand\bigsquare{\mathop{\BigBox}\limits}
\newcommand\MedBox{\vcenter{\hbox{\scalebox{1.2}{$\Box$}}}}

CNNs have achieved overwhelming success in all kinds of machine-learning tasks on visual images. Two key features of CNNs contribute significantly to their success. First, the convolution operation exploits translational symmetry of images by using shared kernels across the whole image. This not only greatly reduces the number of parameters in the network but also allows the parameters to be learned more effectively, as each set of weights will use all locations of the image for learning. Second, CNNs exploit a hierarchical approach \cite{zeiler2014visualizing} for learning image features. The convolution operations can be effectively stacked to form a deep network. Different layers in the CNNs have different receptive fields and therefore can learn features at different scales, with the shallower layers exploiting local neighborhood information and the deeper layers learning more global structures. Such a hierarchical approach proves an effective way to learn images.

Motivated by the success of CNNs, we would like to adopt a similar approach for learning on point (particle) cloud data. However, regular convolution operation cannot be applied on point clouds, as the points there can be distributed irregularly, rather than following some uniform grids as the pixels in an image. Therefore, the basis for a convolution, i.e., a ``local patch'' of each point on which the convolution kernel operates, remains to be defined for point clouds. Moreover, a regular convolution operation, typically in the form $\sum_{j}K_{j}x_{j}$ where $K$ is the kernel and $x_{j}$ denotes the features of each point, is not invariant under permutation of the points. Thus, the form of a convolution also needs to be modified to respect the permutation symmetry of point clouds.

Recently, the edge convolution (``EdgeConv'') operation has been proposed in Ref. \cite{DBLP:journals/corr/abs-1801-07829} as a convolution-like operation for point clouds. EdgeConv starts by representing a point cloud as a graph, whose vertices are the points themselves, and the edges are constructed as connections between each point to its $k$ nearest neighboring points. In this way, a local patch needed for convolution is defined for each point as the $k$ nearest neighboring points connected to it. The EdgeConv operation for each point $x_i$ then has the form
\begin{equation}
\bm{x}_{i}' = \mathop{\bigsquare}_{j=1}^{k} \bm{h}_{\boldsymbol{\Theta}}(\bm{x}_i, \bm{x}_{i_j}),
\label{eq:edge-conv}
\end{equation}
where $\bm{x}_i\in\mathbb{R}^F$ denotes the feature vector of the point $x_i$ and $\{i_1, ..., i_k\}$ are the indices of the $k$ nearest neighboring points of the point $x_i$. The edge function $\bm{h}_{\boldsymbol{\Theta}}: \mathbb{R}^F\times \mathbb{R}^F \rightarrow \mathbb{R}^{F'}$ is some function parametrized by a set of learnable parameters $\boldsymbol{\Theta}$, and $\MedBox$ is a channel-wise symmetric aggregation operation, e.g., $\max$, $\text{sum}$, or $\text{mean}$. The parameters $\boldsymbol{\Theta}$ of the edge function are shared for all points in the point cloud. This, together with the choice of a symmetric aggregation operation $\MedBox$, makes EdgeConv a permutationally symmetric operation on point clouds \footnote{Unlike other approaches in the literature (e.g, Deep Sets \cite{zaheer2017deep}), EdgeConv is not designed as a universal approximator for any permutation-invariant functions. Specifically, the permutation invariance of the EdgeConv layer refers to the fact that the output does not depend on the ordering of the input points. However, to use EdgeConv, one needs to specify a set of features to be used as the ``coordinates'' for the computation of distances needed by the nearest neighbor finding. This choice of a ``coordinate system'' then leads to a canonical ordering of the points in that space where the neighbor relationship is fully determined. Since EdgeConv is performed with the $k$ nearest neighbors for each point, any permutation of the points that changes the neighbor relationship will actually lead to a change in the network output.}.

In this paper, we follow the choice in Ref. \cite{DBLP:journals/corr/abs-1801-07829} to use a specialized form of the edge function,
\begin{equation}
\bm{h}_{\boldsymbol{\Theta}}(\bm{x}_i, \bm{x}_{i_j}) = \bar{\bm{h}}_{\boldsymbol{\Theta}}(\bm{x}_i, \bm{x}_{i_j}-\bm{x}_i),
\label{eq:edge-func}
\end{equation}
where the feature vectors of the neighbors, $\bm{x}_{i_j}$, are substituted by their differences from the central point $\bm{x}_i$ and $\bar{\bm{h}}_{\boldsymbol{\Theta}}$ can be implemented as a  multilayer perceptron (MLP) whose parameters are shared among all edges. For the aggregation operation $\MedBox$, however, we use $\text{mean}$, i.e., $\frac{1}{k}\sum$, throughout this paper, which shows better performance than the $\max$ operation used in the original paper.

One important feature of the EdgeConv operation is that it can be easily stacked, just as regular convolutions. This is because EdgeConv can be viewed as a mapping from a point cloud to another point cloud with the same number of points, only possibly changing the dimension of the feature vector for each point. Therefore, another EdgeConv operation can be applied subsequently. This allows us to build a deep network using EdgeConv operations which can learn features of point clouds hierarchically.

The stackability of EdgeConv operations also brings another interesting possibility. Basically, the feature vectors learned by EdgeConv can be viewed as new coordinates of the original points in a latent space, and then, the distances between points, used in the determination of the $k$ nearest neighbors, can be computed in this latent space. In other words, the proximity of points can be dynamically learned with EdgeConv operations. This results in the DGCNN \cite{DBLP:journals/corr/abs-1801-07829}, in which the graph describing the point clouds are dynamically updated to reflect the changes in the edges, i.e., the neighbors of each point. Reference \cite{DBLP:journals/corr/abs-1801-07829} demonstrates that this leads to better performance than keeping the graph static.

\subsection{ParticleNet}

The ParticleNet architecture makes extensive use of EdgeConv operations and also adopts the dynamic graph update approach. However, a number of different design choices are made in ParticleNet compared to the original DGCNN to better suit the jet tagging task, including the number of neighbors, the configuration of the MLP in EdgeConv, the use of shortcut connection, etc.

\begin{figure}[htbp]
\centering
\includegraphics[height=70mm]{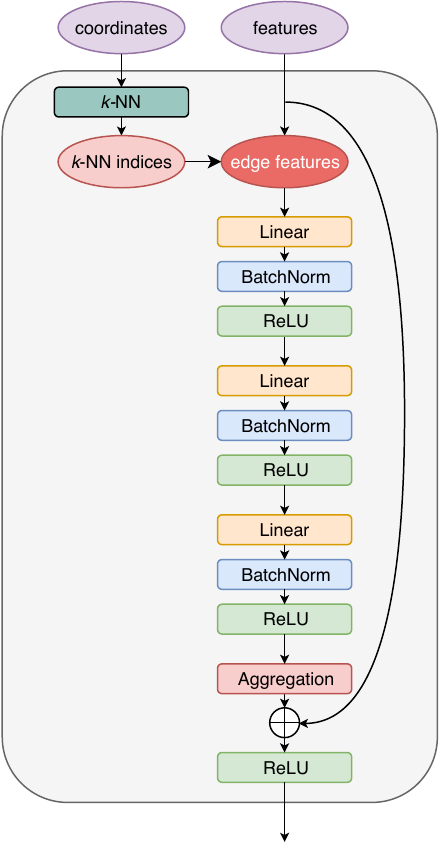}
\caption{The structure of the EdgeConv block.}
\label{fig:block}
\end{figure}

Figure \ref{fig:block} illustrates the structure of the EdgeConv block implemented in this paper. The EdgeConv block starts with finding the $k$ nearest neighboring particles for each particle, using the ``coordinates'' input of the EdgeConv block to compute the distances. Then, inputs to the EdgeConv operation, the ``edge features'', are constructed from the ``features'' input using the indices of $k$ nearest neighboring particles. The EdgeConv operation is implemented as a three-layer MLP. Each layer consists of a linear transformation, followed by a batch normalization \cite{DBLP:journals/corr/IoffeS15} and then the a rectified linear unit (ReLU) \cite{glorot2011deep}. Inspired by ResNet \cite{he2016deep}, a shortcut connection running parallel to the EdgeConv operation is also included in each block, allowing the input features to pass through directly. An EdgeConv block is characterized by two hyperparameters, the number of neighbors $k$, and the number of channels $C=(C_1, C_2, C_3)$, corresponding to the number of units in each linear transformation layer.

\begin{figure}[htbp]
\centering
\subfloat[ParticleNet]{\label{fig:arch}\includegraphics[height=65mm]{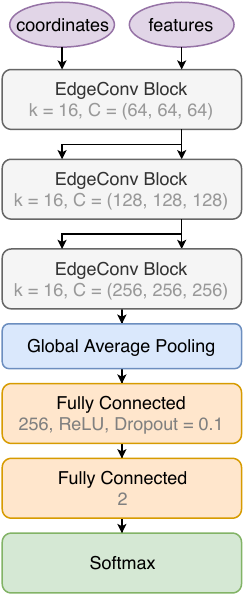}}\hspace{10mm}
\subfloat[ParticleNet-Lite]{\label{fig:arch-lite}\includegraphics[height=65mm]{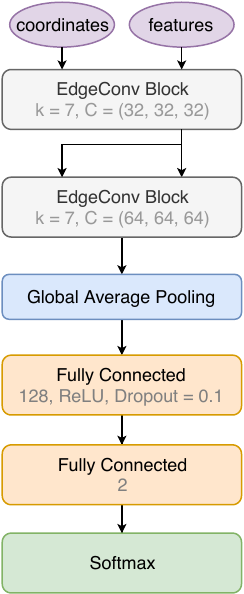}}
\caption{The architectures of the ParticleNet and the ParticleNet-Lite networks.}
\label{fig:networks}
\end{figure}

The ParticleNet architecture used in this paper is shown in Fig. \ref{fig:arch}. It consists of three EdgeConv blocks. The first EdgeConv block uses the spatial coordinates of the particles in the pseudorapidity-azimuth space to compute the distances, while the subsequent blocks use the learned feature vectors as coordinates. The number of nearest neighbors $k$ is 16 for all three blocks, and the number of channels $C$ for each EdgeConv block is (64, 64, 64), (128, 128, 128), and (256, 256, 256), respectively. After the EdgeConv blocks, a channel-wise global average pooling operation is applied to aggregate the learned features over all particles in the cloud. This is followed by a fully connected layer with 256 units and the ReLU activation. A dropout layer \cite{srivastava2014dropout} with a drop probability of 0.1 is included to prevent overfitting. A fully connected layer with two units, followed by a softmax function, is used to generate the output for the binary classification task. 

A similar network with reduced complexity is also investigated. Compared to the baseline ParticleNet architecture, only two EdgeConv blocks are used, with the number of nearest neighbors $k$ reduced to 7 and the number of channels $C$ reduced to (32, 32, 32) and (64, 64, 64) for the two blocks, respectively. The number of units in the fully connected layer after pooling is also lowered to 128. This simplified architecture is denoted as ``ParticleNet-Lite'' and is illustrated in Fig. \ref{fig:arch-lite}. The number of arithmetic operations is reduced by almost an order of magnitude in ParticleNet-Lite, making it more suitable when computational resources are limited.

The networks are implemented with Apache \textsc{MXNet} \cite{DBLP:journals/corr/ChenLLLWWXXZZ15}, and the training is performed on a single Nvidia GTX 1080 Ti graphics card (GPU). A batch size of 384 (1024) is used for the ParticleNet (ParticleNet-Lite) architecture due to GPU memory constraint. The \textsc{AdamW} optimizer \cite{DBLP:journals/corr/abs-1711-05101}, with a weight decay of 0.0001, is used to minimize the cross entropy loss. The one-cycle learning rate (LR) schedule \cite{DBLP:journals/corr/abs-1803-09820} is adopted in the training, with the LR selected following the LR range test described in Ref. \cite{DBLP:journals/corr/abs-1803-09820}, and slightly tuned afterward with a few trial trainings. The training of ParticleNet (ParticleNet-Lite) network uses an initial LR of $3\times10^{-4}$ ($5\times10^{-4}$), rising to the peak LR of $3\times10^{-3}$ ($5\times10^{-3}$) linearly in eight epochs and then decreasing to the initial LR linearly in another eight epochs. This is followed by a cooldown phase of four epochs which gradually reduces the LR to $5\times10^{-7}$ ($1\times10^{-6}$) for better convergence. A snapshot of the model is saved at the end of each epoch, and the model snapshot showing the best accuracy on the validation dataset is selected for the final evaluation.
 \section{Results}
\label{sec:results}

The performance of the ParticleNet architecture is evaluated on two representative jet tagging tasks: top tagging and quark-gluon tagging. In this section, we show the benchmark results.

\begin{table*}[htbp]
\centering
\caption{Input variables used in the top tagging task (TOP) and the quark-gluon tagging task (QG) with and without PID information.
}
\label{tab:input-features}
\resizebox*{\linewidth}{!}{
\begin{ruledtabular}
\begin{tabular}{ccccc}
    Variable & Definition & TOP & QG & QG-PID \\
    \hline
    $\Delta \eta$ & difference in pseudorapidity between the particle and the jet axis & x & x & x \\
    $\Delta \phi$ & difference in azimuthal angle between the particle and the jet axis & x & x & x \\
    \hline
    $\log p_T$ & logarithm of the particle's $p_{T}$ & x & x & x \\
    $\log E$ & logarithm of the particle's energy & x & x & x \\
    $\log \frac{p_T}{p_T(\text{jet})}$ & logarithm of the particle's $p_{T}$ relative to the jet $p_{T}$ & x & x & x \\
    $\log \frac{E}{E(\text{jet})}$ & logarithm of the particle's energy relative to the jet energy & x & x & x \\
    $\Delta R$ & angular separation between the particle and the jet axis ($\sqrt{(\Delta \eta)^2 + (\Delta \phi)^2}$) & x & x & x \\
    \hline
    $q$ & electric charge of the particle &  &  & x \\
    \texttt{isElectron} & if the particle is an electron &  &  & x \\
    \texttt{isMuon} & if the particle is a muon &  &  & x \\
    \texttt{isChargedHadron} & if the particle is a charged hadron &  &  & x \\
    \texttt{isNeutralHadron} & if the particle is a neutral hadron &  &  & x \\
    \texttt{isPhoton} & if the particle is a photon &  &  & x \\
\end{tabular}
\end{ruledtabular}
}
\end{table*}

\subsection{Top tagging}
\label{sec:results-top}

Top tagging, i.e., identifying jets originating from hadronically decaying top quarks, is commonly used in searches for new physics at the LHC. We evaluate the performance of the ParticleNet architecture on this task using the top tagging dataset \cite{kasieczka_gregor_2019_2603256}, which is an extension of the dataset used in Ref. \cite{Butter:2017cot} with some modifications. Jets in this dataset are generated with \textsc{Pythia8} \cite{Sjostrand:2014zea} and passed through \textsc{Delphes} \cite{deFavereau:2013fsa} for fast detector simulation. No multiple parton interaction or pileup is included in the simulation. Jets are clustered from the \textsc{Delphes} E-Flow objects with the anti-$k_{T}$ algorithm \cite{Cacciari:2008gp} using a distance parameter $R=0.8$. Only jets with transverse momentum $p_{T}\in[550, 650]$ and pseudorapidity $|\eta|<2$ are considered. Each signal jet is required to be matched to a hadronically decaying top quark within $\Delta R=0.8$, and all three quarks from the top decay also within $\Delta R=0.8$ of the jet axis. The background jets are obtained from a QCD dijet process. This dataset consists of 2 million jets in total, half signal and half background. The official splitting for training (1.2M jets), validation (400k jets) and testing (400k jets) is used in the development of the ParticleNet model for this dataset. 

In this dataset, up to 200 jet constituent particles are stored for each jet. Only kinematic information, i.e., the 4-momentum $(p_x, p_y, p_z, E)$, of each particle is available. The ParticleNet model takes up to 100 constituent particles with the highest $p_T$ for each jet, and uses seven variables derived from the 4-momentum for each particle as inputs, which are listed in Table \ref{tab:input-features}. The $(\Delta \eta, \Delta \phi)$ variables are used as coordinates to compute the distances between particles in the first EdgeConv block. They are also used together with the other five variables, $\log p_T$, $\log E$, $\log \frac{p_T}{p_T(\text{jet})}$, $\log \frac{E}{E(\text{jet})}$ and $\Delta R$, to form the input feature vector for each particle. 

We compare the performance of ParticleNet with three alternative models \footnote{A comprehensive comparison between a wide range of machine learning approaches on this top tagging dataset is presented in Ref.~\cite{Kasieczka:2019dbj}, where an earlier version of ParticleNet is also included.}:
\begin{itemize}

\item \textbf{ResNeXt-50:} The ResNeXt-50 model is a very deep two-dimensional (2D) CNN using jet images as inputs. The ResNeXt architecture \cite{xie2017aggregated} was proposed for generic image classification, and we modify it slightly for the jet tagging task. The model is trained on the top tagging dataset starting from randomly initialized weights. The implementation details can be found in Appendix \ref{app:arch-resnext50}. Note that the ResNeXt-50 architecture is much deeper and therefore has a much larger capacity than most of the CNN architectures \cite{Cogan:2014oua,deOliveira:2015xxd,Baldi:2016fql,Barnard:2016qma,Dreyer:2018nbf,Kasieczka:2017nvn,Macaluso:2018tck,Komiske:2016rsd,ATL-PHYS-PUB-2017-017,Choi:2018dag} explored for jet tagging so far, so evaluating its performance on jet tagging will shed light on whether architectures for generic image classification are also applicable to jet images.

\item \textbf{P-CNN:} The P-CNN is a 14-layer 1D CNN using particle sequences as inputs. The P-CNN architecture was proposed in the CMS particle-based DNN boosted jet tagger \cite{CMS-DP-2017-049} and showed significant improvement in performance compared to a traditional tagger using boosted decision trees and jet-level observables. The model is also trained on the top tagging dataset from scratch, with the implementation details in Appendix \ref{app:arch-p-cnn}.

\item \textbf{PFN:} The Particle Flow Network (PFN) \cite{Komiske:2018cqr} is a recent architecture for jet tagging which also treats a jet as an unordered set of particles, the same as the particle cloud approach in this paper. However, the network is based on the Deep Sets framework \cite{zaheer2017deep}, which uses global symmetric functions and does not exploit local neighborhood information explicitly as the EdgeConv operation. Since the performance of PFN on this top tagging dataset has already been reported in Ref. \cite{Komiske:2018cqr}, we did not reimplement it but just include the results for comparison.

\end{itemize}

\begin{table*}[htbp]
\centering
\caption{Performance comparison on the top tagging benchmark dataset. The ParticleNet, ParticleNet-Lite, P-CNN and ResNeXt-50 models are trained on the top tagging dataset starting from randomly initialized weights. For each model, the training is repeated for 9 times using different randomly initialized weights. The table shows the result from the median-accuracy training, and the standard deviation of the 9 trainings is quoted as the uncertainty to assess the stability to random weight initialization. Uncertainty on the accuracy and AUC are negligible and therefore omitted. The performance of PFN on this dataset is reported in Ref. \cite{Komiske:2018cqr}, and the uncertainty corresponds to the spread in 10 trainings.}
\label{tab:results-top}
\begin{ruledtabular}
\begin{tabular}{ccccc}
                       & Accuracy & AUC & $1/\varepsilon_{b}\text{ at }\varepsilon_{s}=50\%$ & $1/\varepsilon_{b}\text{ at }\varepsilon_{s}=30\%$ \\
    \hline
    ResNeXt-50         & 0.936     & 0.9837            & $302\pm5$      & $1147\pm58$       \\
    P-CNN              & 0.930     & 0.9803            & $201\pm4$      & $759\pm24$        \\
    PFN                & -         & 0.9819            & $247\pm3$      & $888\pm17$        \\
    ParticleNet-Lite   & 0.937     & 0.9844            & $325\pm5$      & $1262\pm49$       \\
    ParticleNet        & \bf 0.940 & \bf 0.9858        & $\bf 397\pm7$  & $\bf 1615\pm93$   \\
\end{tabular}
\end{ruledtabular}
\end{table*}

\begin{figure}[htbp]
    \centering
    \includegraphics[width=\linewidth]{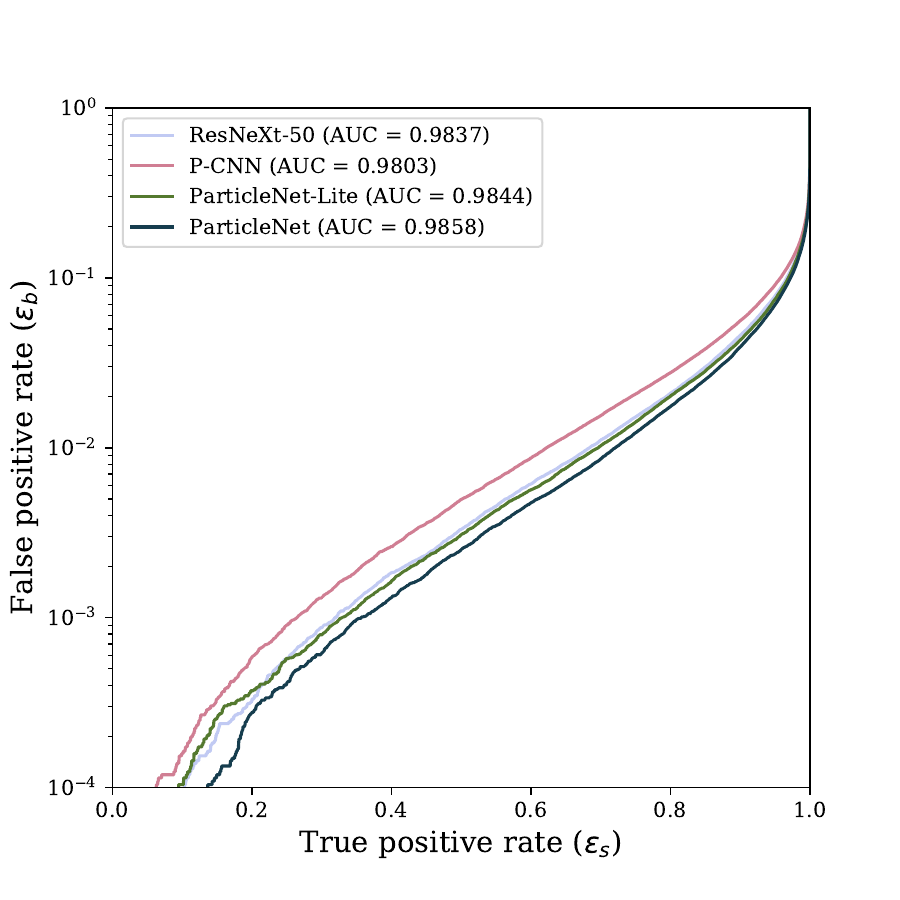}
    \caption{Performance comparison in terms of ROC curves on the top tagging benchmark dataset.}
    \label{fig:roc-top}
\end{figure}

The results are summarized in Table \ref{tab:results-top} and also shown in Fig. \ref{fig:roc-top} in terms of receiver operating characteristic (ROC) curves. A number of metrics are used to evaluate the performance, including the accuracy, the area under the ROC curve (AUC), and the background rejection ($1/\varepsilon_{b}$, i.e., the reciprocal of the background misidentification rate) at a certain signal efficiency ($\varepsilon_{s}$) of 50\% or 30\%. The background rejection metric is particularly relevant to physics analysis at the LHC, as it is directly related to the expected contribution of background, and is commonly used to select the best jet tagging algorithm. The ParticleNet model achieves state-of-the-art performance on the top tagging benchmark dataset and improves over previous methods significantly. Its background rejection power at 30\% signal efficiency is roughly 1.8 (2.1) times as good as PFN (P-CNN), and about 40\% better than ResNeXt-50. Even the ParticleNet-Lite model, with significantly reduced complexity, outperforms all the previous models, achieving about 10\% improvement with respect to ResNeXt-50. The large performance improvement of the ParticleNet architecture over the PFN architecture is likely due to a better exploitation of the local neighborhood information with the EdgeConv operation.

\subsection{Quark-gluon tagging}
\label{sec:results-qg}

Another important jet tagging task is quark-gluon tagging, i.e., discriminating jets initiated by quarks and by gluons. The quark-gluon tagging dataset from Ref. \cite{Komiske:2018cqr} is used to evaluate the performance of the ParticleNet architecture on this task. The signal (quark) and background (gluon) jets are generated with \textsc{Pythia8} using the $Z(\rightarrow\nu\nu)+(u,d,s)$ and $Z(\rightarrow\nu\nu)+g$ processes, respectively. No detector simulation is performed. The final state non-neutrino particles are clustered into jets using the anti-$k_{T}$ algorithm \cite{Cacciari:2008gp} with $R=0.4$. Only jets with transverse momentum $p_{T}\in[500, 550]$ and rapidity $|y|<2$ are considered. This dataset consists of 2 million jets in total, half signal and half background. We follow the recommended splitting of 1.6M/200k/200k for training, validation and testing in the development of the ParticleNet model on this dataset. 

One important difference of the quark-gluon tagging dataset is that it includes not only the four momentum, but also the type of each particle (i.e., electron, photon, pion, etc.). Such particle identification (PID) information can be quite helpful for jet tagging. Therefore, we include this information in the ParticleNet model and compare it with the baseline version using only the kinematic information. The PID information is included in an experimentally realistic way by using only five particle types (electron, muon, charged hadron, neutral hadron, and photon), as well as the electric charge, as inputs. These six additional variables, together with the seven kinematic variables, form the input feature vector of each particle for models with PID information, as shown in Table \ref{tab:input-features}.

\begin{table*}[htbp]
\centering
\caption{Performance comparison on the quark-gluon tagging benchmark dataset. The ParticleNet, ParticleNet-Lite, P-CNN, and ResNeXt-50 models are trained on the quark-gluon tagging dataset starting from randomly initialized weights. The training is repeated 9 times for the ParticleNet model using different randomly initialized weights. The table shows the result from the median-accuracy training, and the standard deviation of the 9 trainings is quoted as the uncertainty to assess the stability to random weight initialization. Because of limited computational resources, the training of other models is performed only once, but the uncertainty due to random weight initialization is expected to be fairly small. The performance of PFN on this dataset is reported in Ref. \cite{Komiske:2018cqr}, and the uncertainty corresponds to the spread in ten trainings. Note that a number of PFN models with different levels of PID information are investigated in Ref. \cite{Komiske:2018cqr}, and ``PFN-Ex'', also using experimentally realistic PID information, is shown here for comparison.}
\label{tab:results-qg}
\begin{ruledtabular}
\begin{tabular}{ccccc}
                & Accuracy & AUC & $1/\varepsilon_{b}\text{ at }\varepsilon_{s}=50\%$ & $1/\varepsilon_{b}\text{ at }\varepsilon_{s}=30\%$ \\
    \hline
    ResNeXt-50           & 0.821     & 0.8960            &  30.9              & 80.8       \\
    P-CNN                & 0.818     & 0.8915            &  31.0              & 82.3       \\
    PFN                  & -         & 0.8911            & $30.8\pm0.4$       & -          \\
    ParticleNet-Lite     & 0.826     & 0.8993            &  32.8              & 84.6       \\
    ParticleNet          & 0.828     & 0.9014            &  33.7              & 85.4       \\
    \hline
    P-CNN (w/ PID)       & 0.827     & 0.9002            &  34.7              & 91.0               \\
    PFN-Ex (w/ PID)      & -         & 0.9005            & $34.7\pm0.4$       & -                  \\
    ParticleNet-Lite (w/ PID) & 0.835 & 0.9079           &  37.1              & 94.5               \\
    ParticleNet (w/ PID) & \bf 0.840 & \bf 0.9116        & $\bf 39.8\pm0.2$   & $\bf 98.6\pm1.3$   \\
\end{tabular}
\end{ruledtabular}
\end{table*}

\begin{figure}[htbp]
    \centering
    \includegraphics[width=\linewidth]{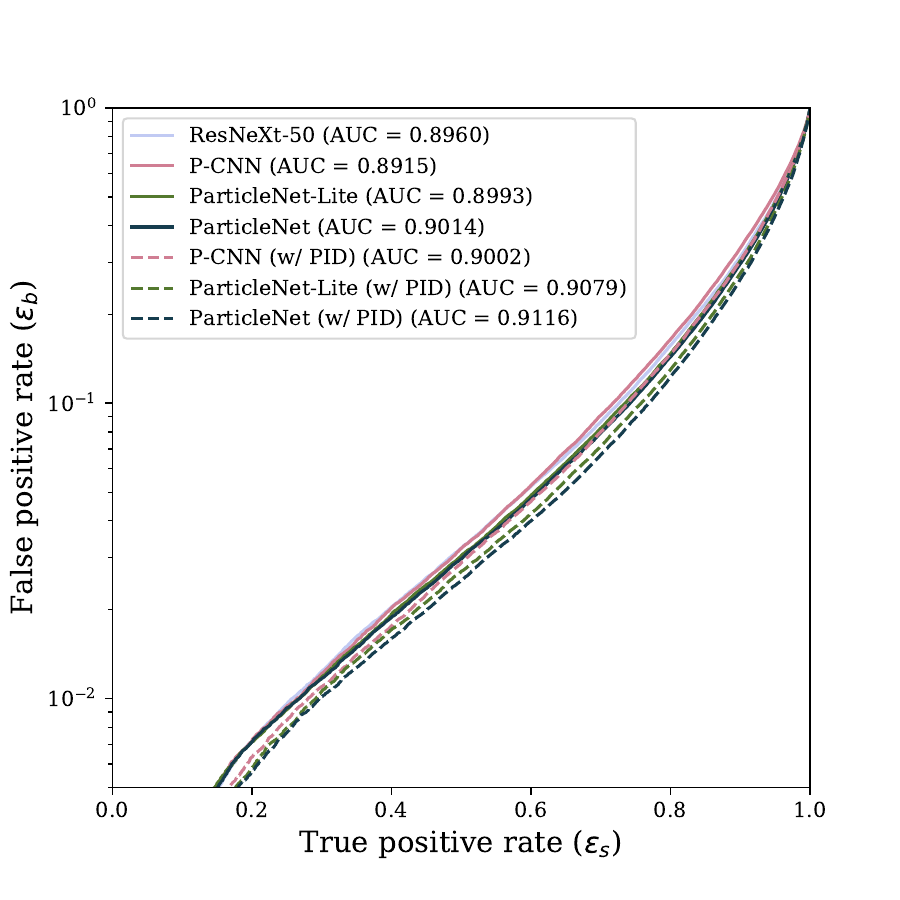}
    \caption{Performance comparison in terms of ROC curves on the quark-gluon tagging benchmark dataset.}
    \label{fig:roc-qg}
\end{figure}

Table \ref{tab:results-qg} compares the performance of the ParticleNet model with a number of alternative models introduced in Sec. \ref{sec:results-top}. Model variants with and without PID inputs are also compared. Note that for the ResNeXt-50 model only the version without PID inputs is presented, as it is based on jet images which cannot incorporate PID information straightforwardly. The corresponding ROC curves are shown in Fig. \ref{fig:roc-qg}. Overall, the addition of PID inputs has a large impact on the performance, increasing the background rejection power by 10\%--15\% compared to the same model without using PID information. This clearly demonstrates the advantage of particle-based jet representations, including the particle cloud representation, as they can easily integrate any additional information for each particle. The best performance is obtained by the ParticleNet model with PID inputs, achieving almost 15\% improvement on the background rejection power compared to the PFN-Ex (PFN using experimentally realistic PID information) and P-CNN models. The ParticleNet-Lite model achieves the second-best performance and shows about 7\% improvement with respect to the PFN-Ex and P-CNN models.

\section{Model complexity}
\label{sec:model-complexity}

Another aspect of machine-learning models is the complexity, e.g., the number of parameters and the computational cost. Table \ref{tab:model-size} compares the number of parameters and the computational cost of all the models used in the top tagging task in Sec. \ref{sec:results-top}. The computational cost is evaluated using the inference time per object, which is a more relevant metric than the training time for real-life applications of machine-learning models. The inference time of each model is measured on both the CPU and the GPU, using the implementations with Apache MXNet. For the CPU, to mimic the  event processing workflow typically used in collider experiments, a batch size of 1 is used, and the inference is performed in single-thread mode. For the GPU, a batch size of 100 is used instead, as the full power of the GPU cannot be revealed with a very small batch size (e.g., 1) due to the overhead in data transfer between the CPU and the GPU. The ParticleNet model achieves the best classification performance at the cost of speed, being more than an order of magnitude slower than the PFN and the P-CNN models, but still it is not prohibitively slow even on the CPU. In addition, the current implementation of the EdgeConv operation used in the ParticleNet model is not as optimized as the regular convolution operation; therefore, further speed-up is expected from an optimized implementation of EdgeConv. On the other hand, the ParticleNet-Lite model provides a good balance between speed and performance, showing more than 40\% improvement in performance while being only a few times slower than the PFN and P-CNN models. Notably, it is also the most economical model, outperforming all previous approaches with only 26k parameters, thanks to the effective exploitation of the permutation symmetry of the particle clouds. Overall, PFN is the fastest model on both the CPU and the GPU, making it a suitable choice for extremely time-critical tasks.

\begin{table*}[htbp]
\centering
\caption{Number of parameters, inference time per object, and background rejection of different models. The CPU inference time is measured on an Intel Core i7-6850K CPU with a single thread using a batch size of 1. The GPU inference time is measured on a Nvidia GTX 1080 Ti GPU using a batch size of 100.}
\label{tab:model-size}
\begin{ruledtabular}
\begin{tabular}{ccccc}
                       & Parameters & Time (CPU) [ms] & Time (GPU) [ms] & $1/\varepsilon_{b}\text{ at }\varepsilon_{s}=30\%$ \\
    \hline
    ResNeXt-50         & 1.46M     & 7.4          & 0.22       & $1147\pm58$       \\
    P-CNN              & 348k      & 1.6          & 0.020      & $759\pm24$        \\
    PFN                & 82k       & \bf 0.8      & \bf 0.018  & $888\pm17$        \\
    ParticleNet-Lite   & \bf 26k   & 2.4          & 0.084      & $1262\pm49$       \\
    ParticleNet        & 366k      & 23           & 0.92       & $\bf 1615\pm93$   \\
\end{tabular}
\end{ruledtabular}
\end{table*}

 \section{Conclusion}
\label{sec:conclusion}

In this paper, we present a new approach for machine learning on jets. The core of this approach is to treat jets as particle clouds, i.e., unordered sets of particles. Based on this particle cloud representation, we introduce ParticleNet, a network architecture tailored to jet tagging tasks. The performance of the ParticleNet architecture is compared with alternative deep-learning architectures, including the jet image--based ResNeXt-50 model, the particle sequence--based P-CNN model and the particle set--based PFN model.  On both the top tagging and the quark-gluon tagging benchmarks, ParticleNet achieves state-of-the-art performance and improves significantly over existing methods. Although the very deep image--based ResNeXt-50 model also shows significant performance improvement over shallower models like P-CNN and PFN on the top-tagging benchmark, indicating that deeper architectures can generally lead to better performance, the gain with the ParticleNet architecture is more substantial. Moreover, the high performance is achieved in a very economical way as the number of trainable parameters is a factor of 4 (56) lower in ParticleNet (ParticleNet-Lite) compared to ResNeXt-50. Such lightweight models are particularly useful for applications in high-energy physics experiments, especially for online event processing in which low latency and memory consumption is critical. 

While we only demonstrate the power of the particle cloud representation in jet tagging tasks, we think that it is a natural and generic way of representing jets (and even the whole collision event) and can be applied to a broad range of particle physics problems. Applications of the particle cloud approach to, e.g., pileup identification, jet grooming, jet energy calibration, etc., would be particularly interesting and worth further investigation.
 
\begin{acknowledgments}
We thank Gregor Kasieczka, Tilman Plehn and Michael Russel for creating the top tagging dataset and sharing it with us, as well as Patrick T. Komiske, Eric M. Metodiev, and Jesse Thaler for creating the quark-gluon tagging dataset and making it publicly accessible. 
This work was supported by the U.S. Department of Energy under Grant No. DE-SC0011702. 
\end{acknowledgments}

\bibliography{cloud-prd}

\begin{thebibliography}{78}%
\makeatletter
\providecommand \@ifxundefined [1]{%
 \@ifx{#1\undefined}
}%
\providecommand \@ifnum [1]{%
 \ifnum #1\expandafter \@firstoftwo
 \else \expandafter \@secondoftwo
 \fi
}%
\providecommand \@ifx [1]{%
 \ifx #1\expandafter \@firstoftwo
 \else \expandafter \@secondoftwo
 \fi
}%
\providecommand \natexlab [1]{#1}%
\providecommand \enquote  [1]{``#1''}%
\providecommand \bibnamefont  [1]{#1}%
\providecommand \bibfnamefont [1]{#1}%
\providecommand \citenamefont [1]{#1}%
\providecommand \href@noop [0]{\@secondoftwo}%
\providecommand \href [0]{\begingroup \@sanitize@url \@href}%
\providecommand \@href[1]{\@@startlink{#1}\@@href}%
\providecommand \@@href[1]{\endgroup#1\@@endlink}%
\providecommand \@sanitize@url [0]{\catcode `\\12\catcode `\$12\catcode
  `\&12\catcode `\#12\catcode `\^12\catcode `\_12\catcode `\%12\relax}%
\providecommand \@@startlink[1]{}%
\providecommand \@@endlink[0]{}%
\providecommand \url  [0]{\begingroup\@sanitize@url \@url }%
\providecommand \@url [1]{\endgroup\@href {#1}{\urlprefix }}%
\providecommand \urlprefix  [0]{URL }%
\providecommand \Eprint [0]{\href }%
\providecommand \doibase [0]{http://dx.doi.org/}%
\providecommand \selectlanguage [0]{\@gobble}%
\providecommand \bibinfo  [0]{\@secondoftwo}%
\providecommand \bibfield  [0]{\@secondoftwo}%
\providecommand \translation [1]{[#1]}%
\providecommand \BibitemOpen [0]{}%
\providecommand \bibitemStop [0]{}%
\providecommand \bibitemNoStop [0]{.\EOS\space}%
\providecommand \EOS [0]{\spacefactor3000\relax}%
\providecommand \BibitemShut  [1]{\csname bibitem#1\endcsname}%
\let\auto@bib@innerbib\@empty
%</preamble>
\bibitem [{\citenamefont {Gallicchio}\ and\ \citenamefont
  {Schwartz}(2011)}]{Gallicchio:2011xq}%
  \BibitemOpen
  \bibfield  {author} {\bibinfo {author} {\bibfnamefont {J.}~\bibnamefont
  {Gallicchio}}\ and\ \bibinfo {author} {\bibfnamefont {M.~D.}\ \bibnamefont
  {Schwartz}},\ }\bibfield  {title} {\enquote {\bibinfo {title} {{Quark and
  Gluon Tagging at the LHC}},}\ }\href {\doibase
  10.1103/PhysRevLett.107.172001} {\bibfield  {journal} {\bibinfo  {journal}
  {Phys. Rev. Lett.}\ }\textbf {\bibinfo {volume} {107}},\ \bibinfo {pages}
  {172001} (\bibinfo {year} {2011})},\ \Eprint {http://arxiv.org/abs/1106.3076}
  {arXiv:1106.3076 [hep-ph]} \BibitemShut {NoStop}%
%%CITATION = ARXIV:1106.3076;%%
\bibitem [{\citenamefont {Gallicchio}\ and\ \citenamefont
  {Schwartz}(2013)}]{Gallicchio:2012ez}%
  \BibitemOpen
  \bibfield  {author} {\bibinfo {author} {\bibfnamefont {J.}~\bibnamefont
  {Gallicchio}}\ and\ \bibinfo {author} {\bibfnamefont {M.~D.}\ \bibnamefont
  {Schwartz}},\ }\bibfield  {title} {\enquote {\bibinfo {title} {{Quark and
  Gluon Jet Substructure}},}\ }\href {\doibase 10.1007/JHEP04(2013)090}
  {\bibfield  {journal} {\bibinfo  {journal} {JHEP}\ }\textbf {\bibinfo
  {volume} {04}},\ \bibinfo {pages} {090} (\bibinfo {year} {2013})},\ \Eprint
  {http://arxiv.org/abs/1211.7038} {arXiv:1211.7038 [hep-ph]} \BibitemShut
  {NoStop}%
%%CITATION = ARXIV:1211.7038;%%
\bibitem [{\citenamefont {Larkoski}\ \emph
  {et~al.}(2014{\natexlab{a}})\citenamefont {Larkoski}, \citenamefont
  {Thaler},\ and\ \citenamefont {Waalewijn}}]{Larkoski:2014pca}%
  \BibitemOpen
  \bibfield  {author} {\bibinfo {author} {\bibfnamefont {A.~J.}\ \bibnamefont
  {Larkoski}}, \bibinfo {author} {\bibfnamefont {J.}~\bibnamefont {Thaler}}, \
  and\ \bibinfo {author} {\bibfnamefont {W.~J.}\ \bibnamefont {Waalewijn}},\
  }\bibfield  {title} {\enquote {\bibinfo {title} {{Gaining (Mutual)
  Information about Quark/Gluon Discrimination}},}\ }\href {\doibase
  10.1007/JHEP11(2014)129} {\bibfield  {journal} {\bibinfo  {journal} {JHEP}\
  }\textbf {\bibinfo {volume} {11}},\ \bibinfo {pages} {129} (\bibinfo {year}
  {2014}{\natexlab{a}})},\ \Eprint {http://arxiv.org/abs/1408.3122}
  {arXiv:1408.3122 [hep-ph]} \BibitemShut {NoStop}%
%%CITATION = ARXIV:1408.3122;%%
\bibitem [{\citenamefont {Bhattacherjee}\ \emph {et~al.}(2015)\citenamefont
  {Bhattacherjee}, \citenamefont {Mukhopadhyay}, \citenamefont {Nojiri},
  \citenamefont {Sakaki},\ and\ \citenamefont
  {Webber}}]{Bhattacherjee:2015psa}%
  \BibitemOpen
  \bibfield  {author} {\bibinfo {author} {\bibfnamefont {B.}~\bibnamefont
  {Bhattacherjee}}, \bibinfo {author} {\bibfnamefont {S.}~\bibnamefont
  {Mukhopadhyay}}, \bibinfo {author} {\bibfnamefont {M.~M.}\ \bibnamefont
  {Nojiri}}, \bibinfo {author} {\bibfnamefont {Y.}~\bibnamefont {Sakaki}}, \
  and\ \bibinfo {author} {\bibfnamefont {B.~R.}\ \bibnamefont {Webber}},\
  }\bibfield  {title} {\enquote {\bibinfo {title} {{Associated jet and subjet
  rates in light-quark and gluon jet discrimination}},}\ }\href {\doibase
  10.1007/JHEP04(2015)131} {\bibfield  {journal} {\bibinfo  {journal} {JHEP}\
  }\textbf {\bibinfo {volume} {04}},\ \bibinfo {pages} {131} (\bibinfo {year}
  {2015})},\ \Eprint {http://arxiv.org/abs/1501.04794} {arXiv:1501.04794
  [hep-ph]} \BibitemShut {NoStop}%
%%CITATION = ARXIV:1501.04794;%%
\bibitem [{\citenamefont {Ferreira~de Lima}\ \emph {et~al.}(2017)\citenamefont
  {Ferreira~de Lima}, \citenamefont {Petrov}, \citenamefont {Soper},\ and\
  \citenamefont {Spannowsky}}]{FerreiradeLima:2016gcz}%
  \BibitemOpen
  \bibfield  {author} {\bibinfo {author} {\bibfnamefont {D.}~\bibnamefont
  {Ferreira~de Lima}}, \bibinfo {author} {\bibfnamefont {P.}~\bibnamefont
  {Petrov}}, \bibinfo {author} {\bibfnamefont {D.}~\bibnamefont {Soper}}, \
  and\ \bibinfo {author} {\bibfnamefont {M.}~\bibnamefont {Spannowsky}},\
  }\bibfield  {title} {\enquote {\bibinfo {title} {{Quark-Gluon tagging with
  Shower Deconstruction: Unearthing dark matter and Higgs couplings}},}\ }\href
  {\doibase 10.1103/PhysRevD.95.034001} {\bibfield  {journal} {\bibinfo
  {journal} {Phys. Rev.}\ }\textbf {\bibinfo {volume} {D95}},\ \bibinfo {pages}
  {034001} (\bibinfo {year} {2017})},\ \Eprint
  {http://arxiv.org/abs/1607.06031} {arXiv:1607.06031 [hep-ph]} \BibitemShut
  {NoStop}%
%%CITATION = ARXIV:1607.06031;%%
\bibitem [{\citenamefont {Gras}\ \emph {et~al.}(2017)\citenamefont {Gras},
  \citenamefont {Höche}, \citenamefont {Kar}, \citenamefont {Larkoski},
  \citenamefont {Lönnblad}, \citenamefont {Plätzer}, \citenamefont
  {Siódmok}, \citenamefont {Skands}, \citenamefont {Soyez},\ and\
  \citenamefont {Thaler}}]{Gras:2017jty}%
  \BibitemOpen
  \bibfield  {author} {\bibinfo {author} {\bibfnamefont {P.}~\bibnamefont
  {Gras}}, \bibinfo {author} {\bibfnamefont {S.}~\bibnamefont {Höche}},
  \bibinfo {author} {\bibfnamefont {D.}~\bibnamefont {Kar}}, \bibinfo {author}
  {\bibfnamefont {A.}~\bibnamefont {Larkoski}}, \bibinfo {author}
  {\bibfnamefont {L.}~\bibnamefont {Lönnblad}}, \bibinfo {author}
  {\bibfnamefont {S.}~\bibnamefont {Plätzer}}, \bibinfo {author}
  {\bibfnamefont {A.}~\bibnamefont {Siódmok}}, \bibinfo {author}
  {\bibfnamefont {P.}~\bibnamefont {Skands}}, \bibinfo {author} {\bibfnamefont
  {G.}~\bibnamefont {Soyez}}, \ and\ \bibinfo {author} {\bibfnamefont
  {J.}~\bibnamefont {Thaler}},\ }\bibfield  {title} {\enquote {\bibinfo {title}
  {{Systematics of quark/gluon tagging}},}\ }\href {\doibase
  10.1007/JHEP07(2017)091} {\bibfield  {journal} {\bibinfo  {journal} {JHEP}\
  }\textbf {\bibinfo {volume} {07}},\ \bibinfo {pages} {091} (\bibinfo {year}
  {2017})},\ \Eprint {http://arxiv.org/abs/1704.03878} {arXiv:1704.03878
  [hep-ph]} \BibitemShut {NoStop}%
%%CITATION = ARXIV:1704.03878;%%
\bibitem [{\citenamefont {Frye}\ \emph {et~al.}(2017)\citenamefont {Frye},
  \citenamefont {Larkoski}, \citenamefont {Thaler},\ and\ \citenamefont
  {Zhou}}]{Frye:2017yrw}%
  \BibitemOpen
  \bibfield  {author} {\bibinfo {author} {\bibfnamefont {C.}~\bibnamefont
  {Frye}}, \bibinfo {author} {\bibfnamefont {A.~J.}\ \bibnamefont {Larkoski}},
  \bibinfo {author} {\bibfnamefont {J.}~\bibnamefont {Thaler}}, \ and\ \bibinfo
  {author} {\bibfnamefont {K.}~\bibnamefont {Zhou}},\ }\bibfield  {title}
  {\enquote {\bibinfo {title} {{Casimir Meets Poisson: Improved Quark/Gluon
  Discrimination with Counting Observables}},}\ }\href {\doibase
  10.1007/JHEP09(2017)083} {\bibfield  {journal} {\bibinfo  {journal} {JHEP}\
  }\textbf {\bibinfo {volume} {09}},\ \bibinfo {pages} {083} (\bibinfo {year}
  {2017})},\ \Eprint {http://arxiv.org/abs/1704.06266} {arXiv:1704.06266
  [hep-ph]} \BibitemShut {NoStop}%
%%CITATION = ARXIV:1704.06266;%%
\bibitem [{\citenamefont {Kaplan}\ \emph {et~al.}(2008)\citenamefont {Kaplan},
  \citenamefont {Rehermann}, \citenamefont {Schwartz},\ and\ \citenamefont
  {Tweedie}}]{Kaplan:2008ie}%
  \BibitemOpen
  \bibfield  {author} {\bibinfo {author} {\bibfnamefont {D.~E.}\ \bibnamefont
  {Kaplan}}, \bibinfo {author} {\bibfnamefont {K.}~\bibnamefont {Rehermann}},
  \bibinfo {author} {\bibfnamefont {M.~D.}\ \bibnamefont {Schwartz}}, \ and\
  \bibinfo {author} {\bibfnamefont {B.}~\bibnamefont {Tweedie}},\ }\bibfield
  {title} {\enquote {\bibinfo {title} {{Top Tagging: A Method for Identifying
  Boosted Hadronically Decaying Top Quarks}},}\ }\href {\doibase
  10.1103/PhysRevLett.101.142001} {\bibfield  {journal} {\bibinfo  {journal}
  {Phys. Rev. Lett.}\ }\textbf {\bibinfo {volume} {101}},\ \bibinfo {pages}
  {142001} (\bibinfo {year} {2008})},\ \Eprint {http://arxiv.org/abs/0806.0848}
  {arXiv:0806.0848 [hep-ph]} \BibitemShut {NoStop}%
%%CITATION = ARXIV:0806.0848;%%
\bibitem [{\citenamefont {Cui}\ \emph {et~al.}(2011)\citenamefont {Cui},
  \citenamefont {Han},\ and\ \citenamefont {Schwartz}}]{Cui:2010km}%
  \BibitemOpen
  \bibfield  {author} {\bibinfo {author} {\bibfnamefont {Y.}~\bibnamefont
  {Cui}}, \bibinfo {author} {\bibfnamefont {Z.}~\bibnamefont {Han}}, \ and\
  \bibinfo {author} {\bibfnamefont {M.~D.}\ \bibnamefont {Schwartz}},\
  }\bibfield  {title} {\enquote {\bibinfo {title} {{W-jet Tagging: Optimizing
  the Identification of Boosted Hadronically-Decaying W Bosons}},}\ }\href
  {\doibase 10.1103/PhysRevD.83.074023} {\bibfield  {journal} {\bibinfo
  {journal} {Phys. Rev.}\ }\textbf {\bibinfo {volume} {D83}},\ \bibinfo {pages}
  {074023} (\bibinfo {year} {2011})},\ \Eprint {http://arxiv.org/abs/1012.2077}
  {arXiv:1012.2077 [hep-ph]} \BibitemShut {NoStop}%
%%CITATION = ARXIV:1012.2077;%%
\bibitem [{\citenamefont {Plehn}\ \emph {et~al.}(2012)\citenamefont {Plehn},
  \citenamefont {Spannowsky},\ and\ \citenamefont {Takeuchi}}]{Plehn:2011sj}%
  \BibitemOpen
  \bibfield  {author} {\bibinfo {author} {\bibfnamefont {T.}~\bibnamefont
  {Plehn}}, \bibinfo {author} {\bibfnamefont {M.}~\bibnamefont {Spannowsky}}, \
  and\ \bibinfo {author} {\bibfnamefont {M.}~\bibnamefont {Takeuchi}},\
  }\bibfield  {title} {\enquote {\bibinfo {title} {{How to Improve Top
  Tagging}},}\ }\href {\doibase 10.1103/PhysRevD.85.034029} {\bibfield
  {journal} {\bibinfo  {journal} {Phys. Rev.}\ }\textbf {\bibinfo {volume}
  {D85}},\ \bibinfo {pages} {034029} (\bibinfo {year} {2012})},\ \Eprint
  {http://arxiv.org/abs/1111.5034} {arXiv:1111.5034 [hep-ph]} \BibitemShut
  {NoStop}%
%%CITATION = ARXIV:1111.5034;%%
\bibitem [{\citenamefont {Soper}\ and\ \citenamefont
  {Spannowsky}(2013)}]{Soper:2012pb}%
  \BibitemOpen
  \bibfield  {author} {\bibinfo {author} {\bibfnamefont {D.~E.}\ \bibnamefont
  {Soper}}\ and\ \bibinfo {author} {\bibfnamefont {M.}~\bibnamefont
  {Spannowsky}},\ }\bibfield  {title} {\enquote {\bibinfo {title} {{Finding top
  quarks with shower deconstruction}},}\ }\href {\doibase
  10.1103/PhysRevD.87.054012} {\bibfield  {journal} {\bibinfo  {journal} {Phys.
  Rev.}\ }\textbf {\bibinfo {volume} {D87}},\ \bibinfo {pages} {054012}
  (\bibinfo {year} {2013})},\ \Eprint {http://arxiv.org/abs/1211.3140}
  {arXiv:1211.3140 [hep-ph]} \BibitemShut {NoStop}%
%%CITATION = ARXIV:1211.3140;%%
\bibitem [{\citenamefont {Anders}\ \emph {et~al.}(2014)\citenamefont {Anders},
  \citenamefont {Bernaciak}, \citenamefont {Kasieczka}, \citenamefont {Plehn},\
  and\ \citenamefont {Schell}}]{Anders:2013oga}%
  \BibitemOpen
  \bibfield  {author} {\bibinfo {author} {\bibfnamefont {C.}~\bibnamefont
  {Anders}}, \bibinfo {author} {\bibfnamefont {C.}~\bibnamefont {Bernaciak}},
  \bibinfo {author} {\bibfnamefont {G.}~\bibnamefont {Kasieczka}}, \bibinfo
  {author} {\bibfnamefont {T.}~\bibnamefont {Plehn}}, \ and\ \bibinfo {author}
  {\bibfnamefont {T.}~\bibnamefont {Schell}},\ }\bibfield  {title} {\enquote
  {\bibinfo {title} {{Benchmarking an even better top tagger algorithm}},}\
  }\href {\doibase 10.1103/PhysRevD.89.074047} {\bibfield  {journal} {\bibinfo
  {journal} {Phys. Rev.}\ }\textbf {\bibinfo {volume} {D89}},\ \bibinfo {pages}
  {074047} (\bibinfo {year} {2014})},\ \Eprint {http://arxiv.org/abs/1312.1504}
  {arXiv:1312.1504 [hep-ph]} \BibitemShut {NoStop}%
%%CITATION = ARXIV:1312.1504;%%
\bibitem [{\citenamefont {Kasieczka}\ \emph {et~al.}(2015)\citenamefont
  {Kasieczka}, \citenamefont {Plehn}, \citenamefont {Schell}, \citenamefont
  {Strebler},\ and\ \citenamefont {Salam}}]{Kasieczka:2015jma}%
  \BibitemOpen
  \bibfield  {author} {\bibinfo {author} {\bibfnamefont {G.}~\bibnamefont
  {Kasieczka}}, \bibinfo {author} {\bibfnamefont {T.}~\bibnamefont {Plehn}},
  \bibinfo {author} {\bibfnamefont {T.}~\bibnamefont {Schell}}, \bibinfo
  {author} {\bibfnamefont {T.}~\bibnamefont {Strebler}}, \ and\ \bibinfo
  {author} {\bibfnamefont {G.~P.}\ \bibnamefont {Salam}},\ }\bibfield  {title}
  {\enquote {\bibinfo {title} {{Resonance Searches with an Updated Top
  Tagger}},}\ }\href {\doibase 10.1007/JHEP06(2015)203} {\bibfield  {journal}
  {\bibinfo  {journal} {JHEP}\ }\textbf {\bibinfo {volume} {06}},\ \bibinfo
  {pages} {203} (\bibinfo {year} {2015})},\ \Eprint
  {http://arxiv.org/abs/1503.05921} {arXiv:1503.05921 [hep-ph]} \BibitemShut
  {NoStop}%
%%CITATION = ARXIV:1503.05921;%%
\bibitem [{\citenamefont {Thaler~{}}\ and\ \citenamefont
  {Van~Tilburg}(2011)}]{Thaler:2010tr}%
  \BibitemOpen
  \bibfield  {author} {\bibinfo {author} {\bibfnamefont {J.}~\bibnamefont
  {Thaler~{}}}\ and\ \bibinfo {author} {\bibfnamefont {K.}~\bibnamefont
  {Van~Tilburg}},\ }\bibfield  {title} {\enquote {\bibinfo {title}
  {{Identifying Boosted Objects with N-subjettiness}},}\ }\href {\doibase
  10.1007/JHEP03(2011)015} {\bibfield  {journal} {\bibinfo  {journal} {JHEP}\
  }\textbf {\bibinfo {volume} {03}},\ \bibinfo {pages} {015} (\bibinfo {year}
  {2011})},\ \Eprint {http://arxiv.org/abs/1011.2268} {arXiv:1011.2268
  [hep-ph]} \BibitemShut {NoStop}%
%%CITATION = ARXIV:1011.2268;%%
\bibitem [{\citenamefont {Thaler~{}}\ and\ \citenamefont
  {Van~Tilburg}(2012)}]{Thaler:2011gf}%
  \BibitemOpen
  \bibfield  {author} {\bibinfo {author} {\bibfnamefont {J.}~\bibnamefont
  {Thaler~{}}}\ and\ \bibinfo {author} {\bibfnamefont {K.}~\bibnamefont
  {Van~Tilburg}},\ }\bibfield  {title} {\enquote {\bibinfo {title} {{Maximizing
  Boosted Top Identification by Minimizing N-subjettiness}},}\ }\href {\doibase
  10.1007/JHEP02(2012)093} {\bibfield  {journal} {\bibinfo  {journal} {JHEP}\
  }\textbf {\bibinfo {volume} {02}},\ \bibinfo {pages} {093} (\bibinfo {year}
  {2012})},\ \Eprint {http://arxiv.org/abs/1108.2701} {arXiv:1108.2701
  [hep-ph]} \BibitemShut {NoStop}%
%%CITATION = ARXIV:1108.2701;%%
\bibitem [{\citenamefont {Larkoski}\ \emph {et~al.}(2013)\citenamefont
  {Larkoski}, \citenamefont {Salam},\ and\ \citenamefont
  {Thaler}}]{Larkoski:2013eya}%
  \BibitemOpen
  \bibfield  {author} {\bibinfo {author} {\bibfnamefont {A.~J.}\ \bibnamefont
  {Larkoski}}, \bibinfo {author} {\bibfnamefont {G.~P.}\ \bibnamefont {Salam}},
  \ and\ \bibinfo {author} {\bibfnamefont {J.}~\bibnamefont {Thaler}},\
  }\bibfield  {title} {\enquote {\bibinfo {title} {{Energy Correlation
  Functions for Jet Substructure}},}\ }\href {\doibase 10.1007/JHEP06(2013)108}
  {\bibfield  {journal} {\bibinfo  {journal} {JHEP}\ }\textbf {\bibinfo
  {volume} {06}},\ \bibinfo {pages} {108} (\bibinfo {year} {2013})},\ \Eprint
  {http://arxiv.org/abs/1305.0007} {arXiv:1305.0007 [hep-ph]} \BibitemShut
  {NoStop}%
%%CITATION = ARXIV:1305.0007;%%
\bibitem [{\citenamefont {Moult}\ \emph {et~al.}(2016)\citenamefont {Moult},
  \citenamefont {Necib},\ and\ \citenamefont {Thaler}}]{Moult:2016cvt}%
  \BibitemOpen
  \bibfield  {author} {\bibinfo {author} {\bibfnamefont {I.}~\bibnamefont
  {Moult}}, \bibinfo {author} {\bibfnamefont {L.}~\bibnamefont {Necib}}, \ and\
  \bibinfo {author} {\bibfnamefont {J.}~\bibnamefont {Thaler}},\ }\bibfield
  {title} {\enquote {\bibinfo {title} {{New Angles on Energy Correlation
  Functions}},}\ }\href {\doibase 10.1007/JHEP12(2016)153} {\bibfield
  {journal} {\bibinfo  {journal} {JHEP}\ }\textbf {\bibinfo {volume} {12}},\
  \bibinfo {pages} {153} (\bibinfo {year} {2016})},\ \Eprint
  {http://arxiv.org/abs/1609.07483} {arXiv:1609.07483 [hep-ph]} \BibitemShut
  {NoStop}%
%%CITATION = ARXIV:1609.07483;%%
\bibitem [{\citenamefont {Larkoski}\ \emph
  {et~al.}(2014{\natexlab{b}})\citenamefont {Larkoski}, \citenamefont
  {Marzani}, \citenamefont {Soyez},\ and\ \citenamefont
  {Thaler}}]{Larkoski:2014wba}%
  \BibitemOpen
  \bibfield  {author} {\bibinfo {author} {\bibfnamefont {A.~J.}\ \bibnamefont
  {Larkoski}}, \bibinfo {author} {\bibfnamefont {S.}~\bibnamefont {Marzani}},
  \bibinfo {author} {\bibfnamefont {G.}~\bibnamefont {Soyez}}, \ and\ \bibinfo
  {author} {\bibfnamefont {J.}~\bibnamefont {Thaler}},\ }\bibfield  {title}
  {\enquote {\bibinfo {title} {{Soft Drop}},}\ }\href {\doibase
  10.1007/JHEP05(2014)146} {\bibfield  {journal} {\bibinfo  {journal} {JHEP}\
  }\textbf {\bibinfo {volume} {05}},\ \bibinfo {pages} {146} (\bibinfo {year}
  {2014}{\natexlab{b}})},\ \Eprint {http://arxiv.org/abs/1402.2657}
  {arXiv:1402.2657 [hep-ph]} \BibitemShut {NoStop}%
%%CITATION = ARXIV:1402.2657;%%
\bibitem [{\citenamefont {Abdesselam}\ \emph {et~al.}(2011)\citenamefont
  {Abdesselam} \emph {et~al.}}]{Abdesselam:2010pt}%
  \BibitemOpen
  \bibfield  {author} {\bibinfo {author} {\bibfnamefont {A.}~\bibnamefont
  {Abdesselam}} \emph {et~al.},\ }\bibfield  {title} {\enquote {\bibinfo
  {title} {{Boosted objects: A Probe of beyond the Standard Model physics}},}\
  }\bibfield  {booktitle} {\emph {\bibinfo {booktitle} {{Boost 2010 Oxford,
  United Kingdom, June 22-25, 2010}}},\ }\href {\doibase
  10.1140/epjc/s10052-011-1661-y} {\bibfield  {journal} {\bibinfo  {journal}
  {Eur. Phys. J.}\ }\textbf {\bibinfo {volume} {C71}},\ \bibinfo {pages} {1661}
  (\bibinfo {year} {2011})},\ \Eprint {http://arxiv.org/abs/1012.5412}
  {arXiv:1012.5412 [hep-ph]} \BibitemShut {NoStop}%
%%CITATION = ARXIV:1012.5412;%%
\bibitem [{\citenamefont {Altheimer}\ \emph {et~al.}(2012)\citenamefont
  {Altheimer} \emph {et~al.}}]{Altheimer:2012mn}%
  \BibitemOpen
  \bibfield  {author} {\bibinfo {author} {\bibfnamefont {A.}~\bibnamefont
  {Altheimer}} \emph {et~al.},\ }\bibfield  {title} {\enquote {\bibinfo {title}
  {{Jet Substructure at the Tevatron and LHC: New results, new tools, new
  benchmarks}},}\ }\bibfield  {booktitle} {\emph {\bibinfo {booktitle} {{BOOST
  2011 Princeton , NJ, USA, 22–26 May 2011}}},\ }\href {\doibase
  10.1088/0954-3899/39/6/063001} {\bibfield  {journal} {\bibinfo  {journal} {J.
  Phys.}\ }\textbf {\bibinfo {volume} {G39}},\ \bibinfo {pages} {063001}
  (\bibinfo {year} {2012})},\ \Eprint {http://arxiv.org/abs/1201.0008}
  {arXiv:1201.0008 [hep-ph]} \BibitemShut {NoStop}%
%%CITATION = ARXIV:1201.0008;%%
\bibitem [{\citenamefont {Altheimer}\ \emph {et~al.}(2014)\citenamefont
  {Altheimer} \emph {et~al.}}]{Altheimer:2013yza}%
  \BibitemOpen
  \bibfield  {author} {\bibinfo {author} {\bibfnamefont {A.}~\bibnamefont
  {Altheimer}} \emph {et~al.},\ }\bibfield  {title} {\enquote {\bibinfo {title}
  {{Boosted objects and jet substructure at the LHC. Report of BOOST2012, held
  at IFIC Valencia, 23rd-27th of July 2012}},}\ }\bibfield  {booktitle} {\emph
  {\bibinfo {booktitle} {{BOOST 2012 Valencia, Spain, July 23-27, 2012}}},\
  }\href {\doibase 10.1140/epjc/s10052-014-2792-8} {\bibfield  {journal}
  {\bibinfo  {journal} {Eur. Phys. J.}\ }\textbf {\bibinfo {volume} {C74}},\
  \bibinfo {pages} {2792} (\bibinfo {year} {2014})},\ \Eprint
  {http://arxiv.org/abs/1311.2708} {arXiv:1311.2708 [hep-ex]} \BibitemShut
  {NoStop}%
%%CITATION = ARXIV:1311.2708;%%
\bibitem [{\citenamefont {Adams}\ \emph {et~al.}(2015)\citenamefont {Adams}
  \emph {et~al.}}]{Adams:2015hiv}%
  \BibitemOpen
  \bibfield  {author} {\bibinfo {author} {\bibfnamefont {D.}~\bibnamefont
  {Adams}} \emph {et~al.},\ }\bibfield  {title} {\enquote {\bibinfo {title}
  {{Towards an Understanding of the Correlations in Jet Substructure}},}\
  }\href {\doibase 10.1140/epjc/s10052-015-3587-2} {\bibfield  {journal}
  {\bibinfo  {journal} {Eur. Phys. J.}\ }\textbf {\bibinfo {volume} {C75}},\
  \bibinfo {pages} {409} (\bibinfo {year} {2015})},\ \Eprint
  {http://arxiv.org/abs/1504.00679} {arXiv:1504.00679 [hep-ph]} \BibitemShut
  {NoStop}%
%%CITATION = ARXIV:1504.00679;%%
\bibitem [{\citenamefont {Larkoski}\ \emph {et~al.}(2020)\citenamefont
  {Larkoski}, \citenamefont {Moult},\ and\ \citenamefont
  {Nachman}}]{Larkoski:2017jix}%
  \BibitemOpen
  \bibfield  {author} {\bibinfo {author} {\bibfnamefont {A.~J.}\ \bibnamefont
  {Larkoski}}, \bibinfo {author} {\bibfnamefont {I.}~\bibnamefont {Moult}}, \
  and\ \bibinfo {author} {\bibfnamefont {B.}~\bibnamefont {Nachman}},\
  }\bibfield  {title} {\enquote {\bibinfo {title} {{Jet Substructure at the
  Large Hadron Collider: A Review of Recent Advances in Theory and Machine
  Learning}},}\ }\href {\doibase 10.1016/j.physrep.2019.11.001} {\bibfield
  {journal} {\bibinfo  {journal} {Phys. Rept.}\ }\textbf {\bibinfo {volume}
  {841}},\ \bibinfo {pages} {1--63} (\bibinfo {year} {2020})},\ \Eprint
  {http://arxiv.org/abs/1709.04464} {arXiv:1709.04464 [hep-ph]} \BibitemShut
  {NoStop}%
%%CITATION = ARXIV:1709.04464;%%
\bibitem [{\citenamefont {Kogler}\ \emph {et~al.}(2019)\citenamefont {Kogler}
  \emph {et~al.}}]{Asquith:2018igt}%
  \BibitemOpen
  \bibfield  {author} {\bibinfo {author} {\bibfnamefont {R.}~\bibnamefont
  {Kogler}} \emph {et~al.},\ }\bibfield  {title} {\enquote {\bibinfo {title}
  {{Jet Substructure at the Large Hadron Collider: Experimental Review}},}\
  }\href {\doibase 10.1103/RevModPhys.91.045003} {\bibfield  {journal}
  {\bibinfo  {journal} {Rev. Mod. Phys.}\ }\textbf {\bibinfo {volume} {91}},\
  \bibinfo {pages} {045003} (\bibinfo {year} {2019})},\ \Eprint
  {http://arxiv.org/abs/1803.06991} {arXiv:1803.06991 [hep-ex]} \BibitemShut
  {NoStop}%
%%CITATION = ARXIV:1803.06991;%%
\bibitem [{\citenamefont {Cogan}\ \emph {et~al.}(2015)\citenamefont {Cogan},
  \citenamefont {Kagan}, \citenamefont {Strauss},\ and\ \citenamefont
  {Schwarztman}}]{Cogan:2014oua}%
  \BibitemOpen
  \bibfield  {author} {\bibinfo {author} {\bibfnamefont {J.}~\bibnamefont
  {Cogan}}, \bibinfo {author} {\bibfnamefont {M.}~\bibnamefont {Kagan}},
  \bibinfo {author} {\bibfnamefont {E.}~\bibnamefont {Strauss}}, \ and\
  \bibinfo {author} {\bibfnamefont {A.}~\bibnamefont {Schwarztman}},\
  }\bibfield  {title} {\enquote {\bibinfo {title} {{Jet-Images: Computer Vision
  Inspired Techniques for Jet Tagging}},}\ }\href {\doibase
  10.1007/JHEP02(2015)118} {\bibfield  {journal} {\bibinfo  {journal} {JHEP}\
  }\textbf {\bibinfo {volume} {02}},\ \bibinfo {pages} {118} (\bibinfo {year}
  {2015})},\ \Eprint {http://arxiv.org/abs/1407.5675} {arXiv:1407.5675
  [hep-ph]} \BibitemShut {NoStop}%
%%CITATION = ARXIV:1407.5675;%%
\bibitem [{\citenamefont {Almeida}\ \emph {et~al.}(2015)\citenamefont
  {Almeida}, \citenamefont {Backović}, \citenamefont {Cliche}, \citenamefont
  {Lee},\ and\ \citenamefont {Perelstein}}]{Almeida:2015jua}%
  \BibitemOpen
  \bibfield  {author} {\bibinfo {author} {\bibfnamefont {L.~G.}\ \bibnamefont
  {Almeida}}, \bibinfo {author} {\bibfnamefont {M.}~\bibnamefont {Backović}},
  \bibinfo {author} {\bibfnamefont {M.}~\bibnamefont {Cliche}}, \bibinfo
  {author} {\bibfnamefont {S.~J.}\ \bibnamefont {Lee}}, \ and\ \bibinfo
  {author} {\bibfnamefont {M.}~\bibnamefont {Perelstein}},\ }\bibfield  {title}
  {\enquote {\bibinfo {title} {{Playing Tag with ANN: Boosted Top
  Identification with Pattern Recognition}},}\ }\href {\doibase
  10.1007/JHEP07(2015)086} {\bibfield  {journal} {\bibinfo  {journal} {JHEP}\
  }\textbf {\bibinfo {volume} {07}},\ \bibinfo {pages} {086} (\bibinfo {year}
  {2015})},\ \Eprint {http://arxiv.org/abs/1501.05968} {arXiv:1501.05968
  [hep-ph]} \BibitemShut {NoStop}%
%%CITATION = ARXIV:1501.05968;%%
\bibitem [{\citenamefont {de~Oliveira}\ \emph {et~al.}(2016)\citenamefont
  {de~Oliveira}, \citenamefont {Kagan}, \citenamefont {Mackey}, \citenamefont
  {Nachman},\ and\ \citenamefont {Schwartzman}}]{deOliveira:2015xxd}%
  \BibitemOpen
  \bibfield  {author} {\bibinfo {author} {\bibfnamefont {L.}~\bibnamefont
  {de~Oliveira}}, \bibinfo {author} {\bibfnamefont {M.}~\bibnamefont {Kagan}},
  \bibinfo {author} {\bibfnamefont {L.}~\bibnamefont {Mackey}}, \bibinfo
  {author} {\bibfnamefont {B.}~\bibnamefont {Nachman}}, \ and\ \bibinfo
  {author} {\bibfnamefont {A.}~\bibnamefont {Schwartzman}},\ }\bibfield
  {title} {\enquote {\bibinfo {title} {{Jet-images — deep learning
  edition}},}\ }\href {\doibase 10.1007/JHEP07(2016)069} {\bibfield  {journal}
  {\bibinfo  {journal} {JHEP}\ }\textbf {\bibinfo {volume} {07}},\ \bibinfo
  {pages} {069} (\bibinfo {year} {2016})},\ \Eprint
  {http://arxiv.org/abs/1511.05190} {arXiv:1511.05190 [hep-ph]} \BibitemShut
  {NoStop}%
%%CITATION = ARXIV:1511.05190;%%
\bibitem [{\citenamefont {Baldi}\ \emph {et~al.}(2016)\citenamefont {Baldi},
  \citenamefont {Bauer}, \citenamefont {Eng}, \citenamefont {Sadowski},\ and\
  \citenamefont {Whiteson}}]{Baldi:2016fql}%
  \BibitemOpen
  \bibfield  {author} {\bibinfo {author} {\bibfnamefont {P.}~\bibnamefont
  {Baldi}}, \bibinfo {author} {\bibfnamefont {K.}~\bibnamefont {Bauer}},
  \bibinfo {author} {\bibfnamefont {C.}~\bibnamefont {Eng}}, \bibinfo {author}
  {\bibfnamefont {P.}~\bibnamefont {Sadowski}}, \ and\ \bibinfo {author}
  {\bibfnamefont {D.}~\bibnamefont {Whiteson}},\ }\bibfield  {title} {\enquote
  {\bibinfo {title} {{Jet Substructure Classification in High-Energy Physics
  with Deep Neural Networks}},}\ }\href {\doibase 10.1103/PhysRevD.93.094034}
  {\bibfield  {journal} {\bibinfo  {journal} {Phys. Rev.}\ }\textbf {\bibinfo
  {volume} {D93}},\ \bibinfo {pages} {094034} (\bibinfo {year} {2016})},\
  \Eprint {http://arxiv.org/abs/1603.09349} {arXiv:1603.09349 [hep-ex]}
  \BibitemShut {NoStop}%
%%CITATION = ARXIV:1603.09349;%%
\bibitem [{\citenamefont {Barnard}\ \emph {et~al.}(2017)\citenamefont
  {Barnard}, \citenamefont {Dawe}, \citenamefont {Dolan},\ and\ \citenamefont
  {Rajcic}}]{Barnard:2016qma}%
  \BibitemOpen
  \bibfield  {author} {\bibinfo {author} {\bibfnamefont {J.}~\bibnamefont
  {Barnard}}, \bibinfo {author} {\bibfnamefont {E.~N.}\ \bibnamefont {Dawe}},
  \bibinfo {author} {\bibfnamefont {M.~J.}\ \bibnamefont {Dolan}}, \ and\
  \bibinfo {author} {\bibfnamefont {N.}~\bibnamefont {Rajcic}},\ }\bibfield
  {title} {\enquote {\bibinfo {title} {{Parton Shower Uncertainties in Jet
  Substructure Analyses with Deep Neural Networks}},}\ }\href {\doibase
  10.1103/PhysRevD.95.014018} {\bibfield  {journal} {\bibinfo  {journal} {Phys.
  Rev.}\ }\textbf {\bibinfo {volume} {D95}},\ \bibinfo {pages} {014018}
  (\bibinfo {year} {2017})},\ \Eprint {http://arxiv.org/abs/1609.00607}
  {arXiv:1609.00607 [hep-ph]} \BibitemShut {NoStop}%
%%CITATION = ARXIV:1609.00607;%%
\bibitem [{\citenamefont {Komiske}\ \emph {et~al.}(2017)\citenamefont
  {Komiske}, \citenamefont {Metodiev},\ and\ \citenamefont
  {Schwartz}}]{Komiske:2016rsd}%
  \BibitemOpen
  \bibfield  {author} {\bibinfo {author} {\bibfnamefont {P.~T.}\ \bibnamefont
  {Komiske}}, \bibinfo {author} {\bibfnamefont {E.~M.}\ \bibnamefont
  {Metodiev}}, \ and\ \bibinfo {author} {\bibfnamefont {M.~D.}\ \bibnamefont
  {Schwartz}},\ }\bibfield  {title} {\enquote {\bibinfo {title} {{Deep learning
  in color: towards automated quark/gluon jet discrimination}},}\ }\href
  {\doibase 10.1007/JHEP01(2017)110} {\bibfield  {journal} {\bibinfo  {journal}
  {JHEP}\ }\textbf {\bibinfo {volume} {01}},\ \bibinfo {pages} {110} (\bibinfo
  {year} {2017})},\ \Eprint {http://arxiv.org/abs/1612.01551} {arXiv:1612.01551
  [hep-ph]} \BibitemShut {NoStop}%
%%CITATION = ARXIV:1612.01551;%%
\bibitem [{\citenamefont {{ATLAS
  Collaboration}}(2017{\natexlab{a}})}]{ATL-PHYS-PUB-2017-017}%
  \BibitemOpen
  \bibfield  {author} {\bibinfo {author} {\bibnamefont {{ATLAS
  Collaboration}}},\ }\href {https://cds.cern.ch/record/2275641} {\emph
  {\bibinfo {title} {{Quark versus Gluon Jet Tagging Using Jet Images with the
  ATLAS Detector}}}},\ \bibinfo {type} {Tech. Rep.}\ \bibinfo {number}
  {ATL-PHYS-PUB-2017-017}\ (\bibinfo  {institution} {CERN},\ \bibinfo {address}
  {Geneva},\ \bibinfo {year} {2017})\BibitemShut {NoStop}%
\bibitem [{\citenamefont {Kasieczka}\ \emph {et~al.}(2017)\citenamefont
  {Kasieczka}, \citenamefont {Plehn}, \citenamefont {Russell},\ and\
  \citenamefont {Schell}}]{Kasieczka:2017nvn}%
  \BibitemOpen
  \bibfield  {author} {\bibinfo {author} {\bibfnamefont {G.}~\bibnamefont
  {Kasieczka}}, \bibinfo {author} {\bibfnamefont {T.}~\bibnamefont {Plehn}},
  \bibinfo {author} {\bibfnamefont {M.}~\bibnamefont {Russell}}, \ and\
  \bibinfo {author} {\bibfnamefont {T.}~\bibnamefont {Schell}},\ }\bibfield
  {title} {\enquote {\bibinfo {title} {{Deep-learning Top Taggers or The End of
  QCD?}}}\ }\href {\doibase 10.1007/JHEP05(2017)006} {\bibfield  {journal}
  {\bibinfo  {journal} {JHEP}\ }\textbf {\bibinfo {volume} {05}},\ \bibinfo
  {pages} {006} (\bibinfo {year} {2017})},\ \Eprint
  {http://arxiv.org/abs/1701.08784} {arXiv:1701.08784 [hep-ph]} \BibitemShut
  {NoStop}%
%%CITATION = ARXIV:1701.08784;%%
\bibitem [{\citenamefont {Macaluso}\ and\ \citenamefont
  {Shih}(2018)}]{Macaluso:2018tck}%
  \BibitemOpen
  \bibfield  {author} {\bibinfo {author} {\bibfnamefont {S.}~\bibnamefont
  {Macaluso}}\ and\ \bibinfo {author} {\bibfnamefont {D.}~\bibnamefont
  {Shih}},\ }\bibfield  {title} {\enquote {\bibinfo {title} {{Pulling Out All
  the Tops with Computer Vision and Deep Learning}},}\ }\href {\doibase
  10.1007/JHEP10(2018)121} {\bibfield  {journal} {\bibinfo  {journal} {JHEP}\
  }\textbf {\bibinfo {volume} {10}},\ \bibinfo {pages} {121} (\bibinfo {year}
  {2018})},\ \Eprint {http://arxiv.org/abs/1803.00107} {arXiv:1803.00107
  [hep-ph]} \BibitemShut {NoStop}%
%%CITATION = ARXIV:1803.00107;%%
\bibitem [{\citenamefont {Choi}\ \emph {et~al.}(2018)\citenamefont {Choi},
  \citenamefont {Lee},\ and\ \citenamefont {Perelstein}}]{Choi:2018dag}%
  \BibitemOpen
  \bibfield  {author} {\bibinfo {author} {\bibfnamefont {S.}~\bibnamefont
  {Choi}}, \bibinfo {author} {\bibfnamefont {S.~J.}\ \bibnamefont {Lee}}, \
  and\ \bibinfo {author} {\bibfnamefont {M.}~\bibnamefont {Perelstein}},\
  }\bibfield  {title} {\enquote {\bibinfo {title} {{Infrared Safety of a
  Neural-Net Top Tagging Algorithm}},}\ }\href@noop {} {\  (\bibinfo {year}
  {2018})},\ \Eprint {http://arxiv.org/abs/1806.01263} {arXiv:1806.01263
  [hep-ph]} \BibitemShut {NoStop}%
%%CITATION = ARXIV:1806.01263;%%
\bibitem [{\citenamefont {Dreyer}\ \emph {et~al.}(2018)\citenamefont {Dreyer},
  \citenamefont {Salam},\ and\ \citenamefont {Soyez}}]{Dreyer:2018nbf}%
  \BibitemOpen
  \bibfield  {author} {\bibinfo {author} {\bibfnamefont {F.~A.}\ \bibnamefont
  {Dreyer}}, \bibinfo {author} {\bibfnamefont {G.~P.}\ \bibnamefont {Salam}}, \
  and\ \bibinfo {author} {\bibfnamefont {G.}~\bibnamefont {Soyez}},\ }\bibfield
   {title} {\enquote {\bibinfo {title} {{The Lund Jet Plane}},}\ }\href
  {\doibase 10.1007/JHEP12(2018)064} {\bibfield  {journal} {\bibinfo  {journal}
  {JHEP}\ }\textbf {\bibinfo {volume} {12}},\ \bibinfo {pages} {064} (\bibinfo
  {year} {2018})},\ \Eprint {http://arxiv.org/abs/1807.04758} {arXiv:1807.04758
  [hep-ph]} \BibitemShut {NoStop}%
%%CITATION = ARXIV:1807.04758;%%
\bibitem [{\citenamefont {Guest}\ \emph {et~al.}(2016)\citenamefont {Guest},
  \citenamefont {Collado}, \citenamefont {Baldi}, \citenamefont {Hsu},
  \citenamefont {Urban},\ and\ \citenamefont {Whiteson}}]{Guest:2016iqz}%
  \BibitemOpen
  \bibfield  {author} {\bibinfo {author} {\bibfnamefont {D.}~\bibnamefont
  {Guest}}, \bibinfo {author} {\bibfnamefont {J.}~\bibnamefont {Collado}},
  \bibinfo {author} {\bibfnamefont {P.}~\bibnamefont {Baldi}}, \bibinfo
  {author} {\bibfnamefont {S.-C.}\ \bibnamefont {Hsu}}, \bibinfo {author}
  {\bibfnamefont {G.}~\bibnamefont {Urban}}, \ and\ \bibinfo {author}
  {\bibfnamefont {D.}~\bibnamefont {Whiteson}},\ }\bibfield  {title} {\enquote
  {\bibinfo {title} {{Jet Flavor Classification in High-Energy Physics with
  Deep Neural Networks}},}\ }\href {\doibase 10.1103/PhysRevD.94.112002}
  {\bibfield  {journal} {\bibinfo  {journal} {Phys. Rev.}\ }\textbf {\bibinfo
  {volume} {D94}},\ \bibinfo {pages} {112002} (\bibinfo {year} {2016})},\
  \Eprint {http://arxiv.org/abs/1607.08633} {arXiv:1607.08633 [hep-ex]}
  \BibitemShut {NoStop}%
%%CITATION = ARXIV:1607.08633;%%
\bibitem [{\citenamefont {Pearkes}\ \emph {et~al.}(2017)\citenamefont
  {Pearkes}, \citenamefont {Fedorko}, \citenamefont {Lister},\ and\
  \citenamefont {Gay}}]{Pearkes:2017hku}%
  \BibitemOpen
  \bibfield  {author} {\bibinfo {author} {\bibfnamefont {J.}~\bibnamefont
  {Pearkes}}, \bibinfo {author} {\bibfnamefont {W.}~\bibnamefont {Fedorko}},
  \bibinfo {author} {\bibfnamefont {A.}~\bibnamefont {Lister}}, \ and\ \bibinfo
  {author} {\bibfnamefont {C.}~\bibnamefont {Gay}},\ }\bibfield  {title}
  {\enquote {\bibinfo {title} {{Jet Constituents for Deep Neural Network Based
  Top Quark Tagging}},}\ }\href@noop {} {\  (\bibinfo {year} {2017})},\ \Eprint
  {http://arxiv.org/abs/1704.02124} {arXiv:1704.02124 [hep-ex]} \BibitemShut
  {NoStop}%
%%CITATION = ARXIV:1704.02124;%%
\bibitem [{\citenamefont {Egan}\ \emph {et~al.}(2017)\citenamefont {Egan},
  \citenamefont {Fedorko}, \citenamefont {Lister}, \citenamefont {Pearkes},\
  and\ \citenamefont {Gay}}]{Egan:2017ojy}%
  \BibitemOpen
  \bibfield  {author} {\bibinfo {author} {\bibfnamefont {S.}~\bibnamefont
  {Egan}}, \bibinfo {author} {\bibfnamefont {W.}~\bibnamefont {Fedorko}},
  \bibinfo {author} {\bibfnamefont {A.}~\bibnamefont {Lister}}, \bibinfo
  {author} {\bibfnamefont {J.}~\bibnamefont {Pearkes}}, \ and\ \bibinfo
  {author} {\bibfnamefont {C.}~\bibnamefont {Gay}},\ }\bibfield  {title}
  {\enquote {\bibinfo {title} {{Long Short-Term Memory (LSTM) networks with jet
  constituents for boosted top tagging at the LHC}},}\ }\href@noop {} {\
  (\bibinfo {year} {2017})},\ \Eprint {http://arxiv.org/abs/1711.09059}
  {arXiv:1711.09059 [hep-ex]} \BibitemShut {NoStop}%
%%CITATION = ARXIV:1711.09059;%%
\bibitem [{\citenamefont {Fraser}\ and\ \citenamefont
  {Schwartz}(2018)}]{Fraser:2018ieu}%
  \BibitemOpen
  \bibfield  {author} {\bibinfo {author} {\bibfnamefont {K.}~\bibnamefont
  {Fraser}}\ and\ \bibinfo {author} {\bibfnamefont {M.~D.}\ \bibnamefont
  {Schwartz}},\ }\bibfield  {title} {\enquote {\bibinfo {title} {{Jet Charge
  and Machine Learning}},}\ }\href {\doibase 10.1007/JHEP10(2018)093}
  {\bibfield  {journal} {\bibinfo  {journal} {JHEP}\ }\textbf {\bibinfo
  {volume} {10}},\ \bibinfo {pages} {093} (\bibinfo {year} {2018})},\ \Eprint
  {http://arxiv.org/abs/1803.08066} {arXiv:1803.08066 [hep-ph]} \BibitemShut
  {NoStop}%
%%CITATION = ARXIV:1803.08066;%%
\bibitem [{\citenamefont {{CMS
  Collaboration}}(2017{\natexlab{a}})}]{CMS-DP-2017-013}%
  \BibitemOpen
  \bibfield  {author} {\bibinfo {author} {\bibnamefont {{CMS Collaboration}}},\
  }\href {https://cds.cern.ch/record/2263802} {\emph {\bibinfo {title} {{CMS
  Phase 1 heavy flavour identification performance and developments}}}},\
  \bibinfo {type} {Tech. Rep.}\ \bibinfo {number} {CMS-DP-2017-013}\ (\bibinfo
  {year} {2017})\BibitemShut {NoStop}%
\bibitem [{\citenamefont {{CMS Collaboration}}(2018)}]{CMS-DP-2018-058}%
  \BibitemOpen
  \bibfield  {author} {\bibinfo {author} {\bibnamefont {{CMS Collaboration}}},\
  }\href {https://cds.cern.ch/record/2646773} {\emph {\bibinfo {title}
  {{Performance of the DeepJet b tagging algorithm using 41.9/fb of data from
  proton-proton collisions at 13TeV with Phase 1 CMS detector}}}},\ \bibinfo
  {type} {Tech. Rep.}\ \bibinfo {number} {CMS-DP-2018-058}\ (\bibinfo {year}
  {2018})\BibitemShut {NoStop}%
\bibitem [{\citenamefont {{CMS
  Collaboration}}(2017{\natexlab{b}})}]{CMS-DP-2017-049}%
  \BibitemOpen
  \bibfield  {author} {\bibinfo {author} {\bibnamefont {{CMS Collaboration}}},\
  }\href {https://cds.cern.ch/record/2295725} {\emph {\bibinfo {title}
  {{Boosted jet identification using particle candidates and deep neural
  networks}}}},\ \bibinfo {type} {Tech. Rep.}\ \bibinfo {number}
  {CMS-DP-2017-049}\ (\bibinfo {year} {2017})\BibitemShut {NoStop}%
\bibitem [{\citenamefont {Stoye}\ \emph {et~al.}(2017)\citenamefont {Stoye},
  \citenamefont {Kieseler}, \citenamefont {Qu}, \citenamefont {Gouskos},
  \citenamefont {Verzetti},\ and\ \citenamefont {Stakia}}]{stoye2017deepjet}%
  \BibitemOpen
  \bibfield  {author} {\bibinfo {author} {\bibfnamefont {M.}~\bibnamefont
  {Stoye}}, \bibinfo {author} {\bibfnamefont {J.}~\bibnamefont {Kieseler}},
  \bibinfo {author} {\bibfnamefont {H.}~\bibnamefont {Qu}}, \bibinfo {author}
  {\bibfnamefont {L.}~\bibnamefont {Gouskos}}, \bibinfo {author} {\bibfnamefont
  {M.}~\bibnamefont {Verzetti}}, \ and\ \bibinfo {author} {\bibfnamefont
  {A.}~\bibnamefont {Stakia}},\ }\bibfield  {title} {\enquote {\bibinfo {title}
  {{DeepJet: Generic physics object based jet multiclass classification for LHC
  experiments}},}\ }in\ \href@noop {} {\emph {\bibinfo {booktitle} {Deep
  Learning for Physical Sciences Workshop at the 31st Conference on Neural
  Information Processing Systems (NIPS)}}}\ (\bibinfo {year}
  {2017})\BibitemShut {NoStop}%
\bibitem [{\citenamefont {{CMS Collaboration}}(2019)}]{CMS-PAS-JME-18-002}%
  \BibitemOpen
  \bibfield  {author} {\bibinfo {author} {\bibnamefont {{CMS Collaboration}}},\
  }\href {https://cds.cern.ch/record/2683870} {\emph {\bibinfo {title}
  {{Machine learning-based identification of highly Lorentz-boosted
  hadronically decaying particles at the CMS experiment}}}},\ \bibinfo {type}
  {Tech. Rep.}\ \bibinfo {number} {CMS-PAS-JME-18-002}\ (\bibinfo
  {institution} {CERN},\ \bibinfo {address} {Geneva},\ \bibinfo {year}
  {2019})\BibitemShut {NoStop}%
\bibitem [{\citenamefont {{ATLAS
  Collaboration}}(2017{\natexlab{b}})}]{ATL-PHYS-PUB-2017-003}%
  \BibitemOpen
  \bibfield  {author} {\bibinfo {author} {\bibnamefont {{ATLAS
  Collaboration}}},\ }\href {https://cds.cern.ch/record/2255226} {\emph
  {\bibinfo {title} {{Identification of Jets Containing $b$-Hadrons with
  Recurrent Neural Networks at the ATLAS Experiment}}}},\ \bibinfo {type}
  {Tech. Rep.}\ \bibinfo {number} {ATL-PHYS-PUB-2017-003}\ (\bibinfo
  {institution} {CERN},\ \bibinfo {address} {Geneva},\ \bibinfo {year}
  {2017})\BibitemShut {NoStop}%
\bibitem [{\citenamefont {Butter}\ \emph {et~al.}(2018)\citenamefont {Butter},
  \citenamefont {Kasieczka}, \citenamefont {Plehn},\ and\ \citenamefont
  {Russell}}]{Butter:2017cot}%
  \BibitemOpen
  \bibfield  {author} {\bibinfo {author} {\bibfnamefont {A.}~\bibnamefont
  {Butter}}, \bibinfo {author} {\bibfnamefont {G.}~\bibnamefont {Kasieczka}},
  \bibinfo {author} {\bibfnamefont {T.}~\bibnamefont {Plehn}}, \ and\ \bibinfo
  {author} {\bibfnamefont {M.}~\bibnamefont {Russell}},\ }\bibfield  {title}
  {\enquote {\bibinfo {title} {{Deep-learned Top Tagging with a Lorentz
  Layer}},}\ }\href {\doibase 10.21468/SciPostPhys.5.3.028} {\bibfield
  {journal} {\bibinfo  {journal} {SciPost Phys.}\ }\textbf {\bibinfo {volume}
  {5}},\ \bibinfo {pages} {028} (\bibinfo {year} {2018})},\ \Eprint
  {http://arxiv.org/abs/1707.08966} {arXiv:1707.08966 [hep-ph]} \BibitemShut
  {NoStop}%
%%CITATION = ARXIV:1707.08966;%%
\bibitem [{\citenamefont {Kasieczka}\ \emph
  {et~al.}(2019{\natexlab{a}})\citenamefont {Kasieczka}, \citenamefont
  {Kiefer}, \citenamefont {Plehn},\ and\ \citenamefont
  {Thompson}}]{Kasieczka:2018lwf}%
  \BibitemOpen
  \bibfield  {author} {\bibinfo {author} {\bibfnamefont {G.}~\bibnamefont
  {Kasieczka}}, \bibinfo {author} {\bibfnamefont {N.}~\bibnamefont {Kiefer}},
  \bibinfo {author} {\bibfnamefont {T.}~\bibnamefont {Plehn}}, \ and\ \bibinfo
  {author} {\bibfnamefont {J.~M.}\ \bibnamefont {Thompson}},\ }\bibfield
  {title} {\enquote {\bibinfo {title} {{Quark-Gluon Tagging: Machine Learning
  vs Detector}},}\ }\href {\doibase 10.21468/SciPostPhys.6.6.069} {\bibfield
  {journal} {\bibinfo  {journal} {SciPost Phys.}\ }\textbf {\bibinfo {volume}
  {6}},\ \bibinfo {pages} {069} (\bibinfo {year} {2019}{\natexlab{a}})},\
  \Eprint {http://arxiv.org/abs/1812.09223} {arXiv:1812.09223 [hep-ph]}
  \BibitemShut {NoStop}%
%%CITATION = ARXIV:1812.09223;%%
\bibitem [{\citenamefont {Erdmann}\ \emph {et~al.}(2018)\citenamefont
  {Erdmann}, \citenamefont {Geiser}, \citenamefont {Rath},\ and\ \citenamefont
  {Rieger}}]{Erdmann:2018shi}%
  \BibitemOpen
  \bibfield  {author} {\bibinfo {author} {\bibfnamefont {M.}~\bibnamefont
  {Erdmann}}, \bibinfo {author} {\bibfnamefont {E.}~\bibnamefont {Geiser}},
  \bibinfo {author} {\bibfnamefont {Y.}~\bibnamefont {Rath}}, \ and\ \bibinfo
  {author} {\bibfnamefont {M.}~\bibnamefont {Rieger}},\ }\bibfield  {title}
  {\enquote {\bibinfo {title} {{Lorentz Boost Networks: Autonomous
  Physics-Inspired Feature Engineering}},}\ }\href@noop {} {\  (\bibinfo {year}
  {2018})},\ \Eprint {http://arxiv.org/abs/1812.09722} {arXiv:1812.09722
  [hep-ex]} \BibitemShut {NoStop}%
%%CITATION = ARXIV:1812.09722;%%
\bibitem [{\citenamefont {Louppe}\ \emph {et~al.}(2019)\citenamefont {Louppe},
  \citenamefont {Cho}, \citenamefont {Becot},\ and\ \citenamefont
  {Cranmer}}]{Louppe:2017ipp}%
  \BibitemOpen
  \bibfield  {author} {\bibinfo {author} {\bibfnamefont {G.}~\bibnamefont
  {Louppe}}, \bibinfo {author} {\bibfnamefont {K.}~\bibnamefont {Cho}},
  \bibinfo {author} {\bibfnamefont {C.}~\bibnamefont {Becot}}, \ and\ \bibinfo
  {author} {\bibfnamefont {K.}~\bibnamefont {Cranmer}},\ }\bibfield  {title}
  {\enquote {\bibinfo {title} {{QCD-Aware Recursive Neural Networks for Jet
  Physics}},}\ }\href {\doibase 10.1007/JHEP01(2019)057} {\bibfield  {journal}
  {\bibinfo  {journal} {JHEP}\ }\textbf {\bibinfo {volume} {01}},\ \bibinfo
  {pages} {057} (\bibinfo {year} {2019})},\ \Eprint
  {http://arxiv.org/abs/1702.00748} {arXiv:1702.00748 [hep-ph]} \BibitemShut
  {NoStop}%
%%CITATION = ARXIV:1702.00748;%%
\bibitem [{\citenamefont {Cheng}(2018)}]{Cheng:2017rdo}%
  \BibitemOpen
  \bibfield  {author} {\bibinfo {author} {\bibfnamefont {T.}~\bibnamefont
  {Cheng}},\ }\bibfield  {title} {\enquote {\bibinfo {title} {{Recursive Neural
  Networks in Quark/Gluon Tagging}},}\ }\href {\doibase
  10.1007/s41781-018-0007-y} {\bibfield  {journal} {\bibinfo  {journal}
  {Comput. Softw. Big Sci.}\ }\textbf {\bibinfo {volume} {2}},\ \bibinfo
  {pages} {3} (\bibinfo {year} {2018})},\ \Eprint
  {http://arxiv.org/abs/1711.02633} {arXiv:1711.02633 [hep-ph]} \BibitemShut
  {NoStop}%
%%CITATION = ARXIV:1711.02633;%%
\bibitem [{\citenamefont {Henrion}\ \emph {et~al.}(2017)\citenamefont
  {Henrion}, \citenamefont {Brehmer}, \citenamefont {Bruna}, \citenamefont
  {Cho}, \citenamefont {Cranmer}, \citenamefont {Louppe},\ and\ \citenamefont
  {Rochette}}]{henrionneural}%
  \BibitemOpen
  \bibfield  {author} {\bibinfo {author} {\bibfnamefont {I.}~\bibnamefont
  {Henrion}}, \bibinfo {author} {\bibfnamefont {J.}~\bibnamefont {Brehmer}},
  \bibinfo {author} {\bibfnamefont {J.}~\bibnamefont {Bruna}}, \bibinfo
  {author} {\bibfnamefont {K.}~\bibnamefont {Cho}}, \bibinfo {author}
  {\bibfnamefont {K.}~\bibnamefont {Cranmer}}, \bibinfo {author} {\bibfnamefont
  {G.}~\bibnamefont {Louppe}}, \ and\ \bibinfo {author} {\bibfnamefont
  {G.}~\bibnamefont {Rochette}},\ }\bibfield  {title} {\enquote {\bibinfo
  {title} {{Neural Message Passing for Jet Physics}},}\ }in\ \href@noop {}
  {\emph {\bibinfo {booktitle} {Deep Learning for Physical Sciences Workshop at
  the 31st Conference on Neural Information Processing Systems (NIPS)}}}\
  (\bibinfo {year} {2017})\BibitemShut {NoStop}%
\bibitem [{\citenamefont {Komiske}\ \emph {et~al.}(2019)\citenamefont
  {Komiske}, \citenamefont {Metodiev},\ and\ \citenamefont
  {Thaler}}]{Komiske:2018cqr}%
  \BibitemOpen
  \bibfield  {author} {\bibinfo {author} {\bibfnamefont {P.~T.}\ \bibnamefont
  {Komiske}}, \bibinfo {author} {\bibfnamefont {E.~M.}\ \bibnamefont
  {Metodiev}}, \ and\ \bibinfo {author} {\bibfnamefont {J.}~\bibnamefont
  {Thaler}},\ }\bibfield  {title} {\enquote {\bibinfo {title} {{Energy Flow
  Networks: Deep Sets for Particle Jets}},}\ }\href {\doibase
  10.1007/JHEP01(2019)121} {\bibfield  {journal} {\bibinfo  {journal} {JHEP}\
  }\textbf {\bibinfo {volume} {01}},\ \bibinfo {pages} {121} (\bibinfo {year}
  {2019})},\ \Eprint {http://arxiv.org/abs/1810.05165} {arXiv:1810.05165
  [hep-ph]} \BibitemShut {NoStop}%
%%CITATION = ARXIV:1810.05165;%%
\bibitem [{\citenamefont {Metodiev}\ \emph {et~al.}(2017)\citenamefont
  {Metodiev}, \citenamefont {Nachman},\ and\ \citenamefont
  {Thaler}}]{Metodiev:2017vrx}%
  \BibitemOpen
  \bibfield  {author} {\bibinfo {author} {\bibfnamefont {E.~M.}\ \bibnamefont
  {Metodiev}}, \bibinfo {author} {\bibfnamefont {B.}~\bibnamefont {Nachman}}, \
  and\ \bibinfo {author} {\bibfnamefont {J.}~\bibnamefont {Thaler}},\
  }\bibfield  {title} {\enquote {\bibinfo {title} {{Classification without
  labels: Learning from mixed samples in high energy physics}},}\ }\href
  {\doibase 10.1007/JHEP10(2017)174} {\bibfield  {journal} {\bibinfo  {journal}
  {JHEP}\ }\textbf {\bibinfo {volume} {10}},\ \bibinfo {pages} {174} (\bibinfo
  {year} {2017})},\ \Eprint {http://arxiv.org/abs/1708.02949} {arXiv:1708.02949
  [hep-ph]} \BibitemShut {NoStop}%
%%CITATION = ARXIV:1708.02949;%%
\bibitem [{\citenamefont {Komiske}\ \emph
  {et~al.}(2018{\natexlab{a}})\citenamefont {Komiske}, \citenamefont
  {Metodiev}, \citenamefont {Nachman},\ and\ \citenamefont
  {Schwartz}}]{Komiske:2018oaa}%
  \BibitemOpen
  \bibfield  {author} {\bibinfo {author} {\bibfnamefont {P.~T.}\ \bibnamefont
  {Komiske}}, \bibinfo {author} {\bibfnamefont {E.~M.}\ \bibnamefont
  {Metodiev}}, \bibinfo {author} {\bibfnamefont {B.}~\bibnamefont {Nachman}}, \
  and\ \bibinfo {author} {\bibfnamefont {M.~D.}\ \bibnamefont {Schwartz}},\
  }\bibfield  {title} {\enquote {\bibinfo {title} {{Learning to classify from
  impure samples with high-dimensional data}},}\ }\href {\doibase
  10.1103/PhysRevD.98.011502} {\bibfield  {journal} {\bibinfo  {journal} {Phys.
  Rev.}\ }\textbf {\bibinfo {volume} {D98}},\ \bibinfo {pages} {011502(R)}
  (\bibinfo {year} {2018}{\natexlab{a}})},\ \Eprint
  {http://arxiv.org/abs/1801.10158} {arXiv:1801.10158 [hep-ph]} \BibitemShut
  {NoStop}%
%%CITATION = ARXIV:1801.10158;%%
\bibitem [{\citenamefont {Andreassen}\ \emph {et~al.}(2019)\citenamefont
  {Andreassen}, \citenamefont {Feige}, \citenamefont {Frye},\ and\
  \citenamefont {Schwartz}}]{Andreassen:2018apy}%
  \BibitemOpen
  \bibfield  {author} {\bibinfo {author} {\bibfnamefont {A.}~\bibnamefont
  {Andreassen}}, \bibinfo {author} {\bibfnamefont {I.}~\bibnamefont {Feige}},
  \bibinfo {author} {\bibfnamefont {C.}~\bibnamefont {Frye}}, \ and\ \bibinfo
  {author} {\bibfnamefont {M.~D.}\ \bibnamefont {Schwartz}},\ }\bibfield
  {title} {\enquote {\bibinfo {title} {{JUNIPR: a Framework for Unsupervised
  Machine Learning in Particle Physics}},}\ }\href {\doibase
  10.1140/epjc/s10052-019-6607-9} {\bibfield  {journal} {\bibinfo  {journal}
  {Eur. Phys. J.}\ }\textbf {\bibinfo {volume} {C79}},\ \bibinfo {pages} {102}
  (\bibinfo {year} {2019})},\ \Eprint {http://arxiv.org/abs/1804.09720}
  {arXiv:1804.09720 [hep-ph]} \BibitemShut {NoStop}%
%%CITATION = ARXIV:1804.09720;%%
\bibitem [{\citenamefont {Komiske}\ \emph
  {et~al.}(2018{\natexlab{b}})\citenamefont {Komiske}, \citenamefont
  {Metodiev},\ and\ \citenamefont {Thaler}}]{Komiske:2018vkc}%
  \BibitemOpen
  \bibfield  {author} {\bibinfo {author} {\bibfnamefont {P.~T.}\ \bibnamefont
  {Komiske}}, \bibinfo {author} {\bibfnamefont {E.~M.}\ \bibnamefont
  {Metodiev}}, \ and\ \bibinfo {author} {\bibfnamefont {J.}~\bibnamefont
  {Thaler}},\ }\bibfield  {title} {\enquote {\bibinfo {title} {{An operational
  definition of quark and gluon jets}},}\ }\href {\doibase
  10.1007/JHEP11(2018)059} {\bibfield  {journal} {\bibinfo  {journal} {JHEP}\
  }\textbf {\bibinfo {volume} {11}},\ \bibinfo {pages} {059} (\bibinfo {year}
  {2018}{\natexlab{b}})},\ \Eprint {http://arxiv.org/abs/1809.01140}
  {arXiv:1809.01140 [hep-ph]} \BibitemShut {NoStop}%
%%CITATION = ARXIV:1809.01140;%%
\bibitem [{Note1()}]{Note1}%
  \BibitemOpen
  \bibinfo {note} {The idea of regarding jets as unordered sets of particles
  was also proposed in Ref. \cite {Komiske:2018cqr} independently while this
  work was being finalized. We provide comparison to their approach in later
  sections.}\BibitemShut {Stop}%
\bibitem [{\citenamefont {Wang}\ \emph {et~al.}(2019)\citenamefont {Wang},
  \citenamefont {Sun}, \citenamefont {Liu}, \citenamefont {Sarma},
  \citenamefont {Bronstein},\ and\ \citenamefont
  {Solomon}}]{DBLP:journals/corr/abs-1801-07829}%
  \BibitemOpen
  \bibfield  {author} {\bibinfo {author} {\bibfnamefont {Y.}~\bibnamefont
  {Wang}}, \bibinfo {author} {\bibfnamefont {Y.}~\bibnamefont {Sun}}, \bibinfo
  {author} {\bibfnamefont {Z.}~\bibnamefont {Liu}}, \bibinfo {author}
  {\bibfnamefont {S.~E.}\ \bibnamefont {Sarma}}, \bibinfo {author}
  {\bibfnamefont {M.~M.}\ \bibnamefont {Bronstein}}, \ and\ \bibinfo {author}
  {\bibfnamefont {J.~M.}\ \bibnamefont {Solomon}},\ }\bibfield  {title}
  {\enquote {\bibinfo {title} {Dynamic graph cnn for learning on point
  clouds},}\ }\href {\doibase 10.1145/3326362} {\bibfield  {journal} {\bibinfo
  {journal} {ACM Trans. Graph.}\ }\textbf {\bibinfo {volume} {38}},\ \bibinfo
  {pages} {146} (\bibinfo {year} {2019})}\BibitemShut {NoStop}%
\bibitem [{\citenamefont {Sirunyan}\ \emph {et~al.}(2017)\citenamefont
  {Sirunyan} \emph {et~al.}}]{Sirunyan:2017ulk}%
  \BibitemOpen
  \bibfield  {author} {\bibinfo {author} {\bibfnamefont {A.~M.}\ \bibnamefont
  {Sirunyan}} \emph {et~al.} (\bibinfo {collaboration} {CMS}),\ }\bibfield
  {title} {\enquote {\bibinfo {title} {{Particle-flow reconstruction and global
  event description with the CMS detector}},}\ }\href {\doibase
  10.1088/1748-0221/12/10/P10003} {\bibfield  {journal} {\bibinfo  {journal}
  {JINST}\ }\textbf {\bibinfo {volume} {12}},\ \bibinfo {pages} {P10003}
  (\bibinfo {year} {2017})},\ \Eprint {http://arxiv.org/abs/1706.04965}
  {arXiv:1706.04965 [physics.ins-det]} \BibitemShut {NoStop}%
%%CITATION = ARXIV:1706.04965;%%
\bibitem [{\citenamefont {Aaboud}\ \emph {et~al.}(2017)\citenamefont {Aaboud}
  \emph {et~al.}}]{Aaboud:2017aca}%
  \BibitemOpen
  \bibfield  {author} {\bibinfo {author} {\bibfnamefont {M.}~\bibnamefont
  {Aaboud}} \emph {et~al.} (\bibinfo {collaboration} {ATLAS}),\ }\bibfield
  {title} {\enquote {\bibinfo {title} {{Jet reconstruction and performance
  using particle flow with the ATLAS Detector}},}\ }\href {\doibase
  10.1140/epjc/s10052-017-5031-2} {\bibfield  {journal} {\bibinfo  {journal}
  {Eur. Phys. J.}\ }\textbf {\bibinfo {volume} {C77}},\ \bibinfo {pages} {466}
  (\bibinfo {year} {2017})},\ \Eprint {http://arxiv.org/abs/1703.10485}
  {arXiv:1703.10485 [hep-ex]} \BibitemShut {NoStop}%
%%CITATION = ARXIV:1703.10485;%%
\bibitem [{\citenamefont {{He}}\ \emph {et~al.}(2016)\citenamefont {{He}},
  \citenamefont {{Zhang}}, \citenamefont {{Ren}},\ and\ \citenamefont
  {{Sun}}}]{he2016deep}%
  \BibitemOpen
  \bibfield  {author} {\bibinfo {author} {\bibfnamefont {K.}~\bibnamefont
  {{He}}}, \bibinfo {author} {\bibfnamefont {X.}~\bibnamefont {{Zhang}}},
  \bibinfo {author} {\bibfnamefont {S.}~\bibnamefont {{Ren}}}, \ and\ \bibinfo
  {author} {\bibfnamefont {J.}~\bibnamefont {{Sun}}},\ }\bibfield  {title}
  {\enquote {\bibinfo {title} {Deep residual learning for image recognition},}\
  }in\ \href {\doibase 10.1109/CVPR.2016.90} {\emph {\bibinfo {booktitle} {2016
  IEEE Conference on Computer Vision and Pattern Recognition (CVPR)}}}\
  (\bibinfo  {publisher} {IEEE},\ \bibinfo {address} {Las Vegas, NV, USA},\
  \bibinfo {year} {2016})\ pp.\ \bibinfo {pages} {770--778}\BibitemShut
  {NoStop}%
\bibitem [{\citenamefont {{Szegedy}}\ \emph {et~al.}(2016)\citenamefont
  {{Szegedy}}, \citenamefont {{Vanhoucke}}, \citenamefont {{Ioffe}},
  \citenamefont {{Shlens}},\ and\ \citenamefont
  {{Wojna}}}]{szegedy2016rethinking}%
  \BibitemOpen
  \bibfield  {author} {\bibinfo {author} {\bibfnamefont {C.}~\bibnamefont
  {{Szegedy}}}, \bibinfo {author} {\bibfnamefont {V.}~\bibnamefont
  {{Vanhoucke}}}, \bibinfo {author} {\bibfnamefont {S.}~\bibnamefont
  {{Ioffe}}}, \bibinfo {author} {\bibfnamefont {J.}~\bibnamefont {{Shlens}}}, \
  and\ \bibinfo {author} {\bibfnamefont {Z.}~\bibnamefont {{Wojna}}},\
  }\bibfield  {title} {\enquote {\bibinfo {title} {Rethinking the inception
  architecture for computer vision},}\ }in\ \href {\doibase
  10.1109/CVPR.2016.308} {\emph {\bibinfo {booktitle} {2016 IEEE Conference on
  Computer Vision and Pattern Recognition (CVPR)}}}\ (\bibinfo  {publisher}
  {IEEE},\ \bibinfo {address} {Las Vegas, NV, USA},\ \bibinfo {year} {2016})\
  pp.\ \bibinfo {pages} {2818--2826}\BibitemShut {NoStop}%
\bibitem [{\citenamefont {Zaheer}\ \emph {et~al.}(2017)\citenamefont {Zaheer},
  \citenamefont {Kottur}, \citenamefont {Ravanbakhsh}, \citenamefont {Poczos},
  \citenamefont {Salakhutdinov},\ and\ \citenamefont {Smola}}]{zaheer2017deep}%
  \BibitemOpen
  \bibfield  {author} {\bibinfo {author} {\bibfnamefont {M.}~\bibnamefont
  {Zaheer}}, \bibinfo {author} {\bibfnamefont {S.}~\bibnamefont {Kottur}},
  \bibinfo {author} {\bibfnamefont {S.}~\bibnamefont {Ravanbakhsh}}, \bibinfo
  {author} {\bibfnamefont {B.}~\bibnamefont {Poczos}}, \bibinfo {author}
  {\bibfnamefont {R.~R.}\ \bibnamefont {Salakhutdinov}}, \ and\ \bibinfo
  {author} {\bibfnamefont {A.~J.}\ \bibnamefont {Smola}},\ }\bibfield  {title}
  {\enquote {\bibinfo {title} {Deep sets},}\ }in\ \href
  {http://papers.nips.cc/paper/6931-deep-sets.pdf} {\emph {\bibinfo {booktitle}
  {Advances in Neural Information Processing Systems 30}}}\ (\bibinfo
  {publisher} {Curran Associates, Inc.},\ \bibinfo {year} {2017})\ pp.\
  \bibinfo {pages} {3391--3401}\BibitemShut {NoStop}%
\bibitem [{\citenamefont {Zeiler}\ and\ \citenamefont
  {Fergus}(2014)}]{zeiler2014visualizing}%
  \BibitemOpen
  \bibfield  {author} {\bibinfo {author} {\bibfnamefont {M.~D.}\ \bibnamefont
  {Zeiler}}\ and\ \bibinfo {author} {\bibfnamefont {R.}~\bibnamefont
  {Fergus}},\ }\bibfield  {title} {\enquote {\bibinfo {title} {Visualizing and
  understanding convolutional networks},}\ }in\ \href {\doibase
  10.1007/978-3-319-10590-1_53} {\emph {\bibinfo {booktitle} {Computer Vision
  -- ECCV 2014}}}\ (\bibinfo  {publisher} {Springer},\ \bibinfo {address}
  {Cham},\ \bibinfo {year} {2014})\ pp.\ \bibinfo {pages}
  {818--833}\BibitemShut {NoStop}%
\bibitem [{Note2()}]{Note2}%
  \BibitemOpen
  \bibinfo {note} {Unlike other approaches in the literature (e.g, Deep Sets
  \cite {zaheer2017deep}), EdgeConv is not designed as a universal approximator
  for any permutation-invariant functions. Specifically, the permutation
  invariance of the EdgeConv layer refers to the fact that the output does not
  depend on the ordering of the input points. However, to use EdgeConv, one
  needs to specify a set of features to be used as the ``coordinates'' for the
  computation of distances needed by the nearest neighbor finding. This choice
  of a ``coordinate system'' then leads to a canonical ordering of the points
  in that space where the neighbor relationship is fully determined. Since
  EdgeConv is performed with the $k$ nearest neighbors for each point, any
  permutation of the points that changes the neighbor relationship will
  actually lead to a change in the network output.}\BibitemShut {Stop}%
\bibitem [{\citenamefont {Ioffe}\ and\ \citenamefont
  {Szegedy}(2015)}]{DBLP:journals/corr/IoffeS15}%
  \BibitemOpen
  \bibfield  {author} {\bibinfo {author} {\bibfnamefont {S.}~\bibnamefont
  {Ioffe}}\ and\ \bibinfo {author} {\bibfnamefont {C.}~\bibnamefont
  {Szegedy}},\ }\bibfield  {title} {\enquote {\bibinfo {title} {Batch
  normalization: Accelerating deep network training by reducing internal
  covariate shift},}\ }in\ \href
  {http://proceedings.mlr.press/v37/ioffe15.html} {\emph {\bibinfo {booktitle}
  {Proceedings of the 32nd International Conference on Machine Learning}}},\
  Vol.~\bibinfo {volume} {37}\ (\bibinfo  {publisher} {PMLR},\ \bibinfo
  {address} {Lille, France},\ \bibinfo {year} {2015})\ pp.\ \bibinfo {pages}
  {448--456}\BibitemShut {NoStop}%
\bibitem [{\citenamefont {Glorot}\ \emph {et~al.}(2011)\citenamefont {Glorot},
  \citenamefont {Bordes},\ and\ \citenamefont {Bengio}}]{glorot2011deep}%
  \BibitemOpen
  \bibfield  {author} {\bibinfo {author} {\bibfnamefont {X.}~\bibnamefont
  {Glorot}}, \bibinfo {author} {\bibfnamefont {A.}~\bibnamefont {Bordes}}, \
  and\ \bibinfo {author} {\bibfnamefont {Y.}~\bibnamefont {Bengio}},\
  }\bibfield  {title} {\enquote {\bibinfo {title} {Deep sparse rectifier neural
  networks},}\ }in\ \href {http://proceedings.mlr.press/v15/glorot11a.html}
  {\emph {\bibinfo {booktitle} {Proceedings of the Fourteenth International
  Conference on Artificial Intelligence and Statistics}}},\ Vol.~\bibinfo
  {volume} {15}\ (\bibinfo  {publisher} {PMLR},\ \bibinfo {address} {Fort
  Lauderdale, FL, USA},\ \bibinfo {year} {2011})\ pp.\ \bibinfo {pages}
  {315--323}\BibitemShut {NoStop}%
\bibitem [{\citenamefont {Srivastava}\ \emph {et~al.}(2014)\citenamefont
  {Srivastava}, \citenamefont {Hinton}, \citenamefont {Krizhevsky},
  \citenamefont {Sutskever},\ and\ \citenamefont
  {Salakhutdinov}}]{srivastava2014dropout}%
  \BibitemOpen
  \bibfield  {author} {\bibinfo {author} {\bibfnamefont {N.}~\bibnamefont
  {Srivastava}}, \bibinfo {author} {\bibfnamefont {G.}~\bibnamefont {Hinton}},
  \bibinfo {author} {\bibfnamefont {A.}~\bibnamefont {Krizhevsky}}, \bibinfo
  {author} {\bibfnamefont {I.}~\bibnamefont {Sutskever}}, \ and\ \bibinfo
  {author} {\bibfnamefont {R.}~\bibnamefont {Salakhutdinov}},\ }\bibfield
  {title} {\enquote {\bibinfo {title} {Dropout: A simple way to prevent neural
  networks from overfitting},}\ }\href
  {http://jmlr.org/papers/v15/srivastava14a.html} {\bibfield  {journal}
  {\bibinfo  {journal} {Journal of Machine Learning Research}\ }\textbf
  {\bibinfo {volume} {15}},\ \bibinfo {pages} {1929--1958} (\bibinfo {year}
  {2014})}\BibitemShut {NoStop}%
\bibitem [{\citenamefont {Chen}\ \emph {et~al.}(2015)\citenamefont {Chen},
  \citenamefont {Li}, \citenamefont {Li}, \citenamefont {Lin}, \citenamefont
  {Wang}, \citenamefont {Wang}, \citenamefont {Xiao}, \citenamefont {Xu},
  \citenamefont {Zhang},\ and\ \citenamefont
  {Zhang}}]{DBLP:journals/corr/ChenLLLWWXXZZ15}%
  \BibitemOpen
  \bibfield  {author} {\bibinfo {author} {\bibfnamefont {T.}~\bibnamefont
  {Chen}}, \bibinfo {author} {\bibfnamefont {M.}~\bibnamefont {Li}}, \bibinfo
  {author} {\bibfnamefont {Y.}~\bibnamefont {Li}}, \bibinfo {author}
  {\bibfnamefont {M.}~\bibnamefont {Lin}}, \bibinfo {author} {\bibfnamefont
  {N.}~\bibnamefont {Wang}}, \bibinfo {author} {\bibfnamefont {M.}~\bibnamefont
  {Wang}}, \bibinfo {author} {\bibfnamefont {T.}~\bibnamefont {Xiao}}, \bibinfo
  {author} {\bibfnamefont {B.}~\bibnamefont {Xu}}, \bibinfo {author}
  {\bibfnamefont {C.}~\bibnamefont {Zhang}}, \ and\ \bibinfo {author}
  {\bibfnamefont {Z.}~\bibnamefont {Zhang}},\ }\bibfield  {title} {\enquote
  {\bibinfo {title} {{MXNet: {A} Flexible and Efficient Machine Learning
  Library for Heterogeneous Distributed Systems}},}\ }in\ \href@noop {} {\emph
  {\bibinfo {booktitle} {Workshop on Machine Learning Systems at the 29st
  Conference on Neural Information Processing Systems (NIPS)}}}\ (\bibinfo
  {publisher} {LearningSys.org},\ \bibinfo {address} {Montreal, Canada},\
  \bibinfo {year} {2015})\ \Eprint {http://arxiv.org/abs/1512.01274}
  {arXiv:1512.01274} \BibitemShut {NoStop}%
\bibitem [{\citenamefont {Loshchilov}\ and\ \citenamefont
  {Hutter}(2017)}]{DBLP:journals/corr/abs-1711-05101}%
  \BibitemOpen
  \bibfield  {author} {\bibinfo {author} {\bibfnamefont {I.}~\bibnamefont
  {Loshchilov}}\ and\ \bibinfo {author} {\bibfnamefont {F.}~\bibnamefont
  {Hutter}},\ }\bibfield  {title} {\enquote {\bibinfo {title} {Fixing weight
  decay regularization in adam},}\ }\href@noop {} {\  (\bibinfo {year}
  {2017})},\ \Eprint {http://arxiv.org/abs/1711.05101} {arXiv:1711.05101}
  \BibitemShut {NoStop}%
\bibitem [{\citenamefont {Smith}(2018)}]{DBLP:journals/corr/abs-1803-09820}%
  \BibitemOpen
  \bibfield  {author} {\bibinfo {author} {\bibfnamefont {L.~N.}\ \bibnamefont
  {Smith}},\ }\bibfield  {title} {\enquote {\bibinfo {title} {A disciplined
  approach to neural network hyper-parameters: Part 1 - learning rate, batch
  size, momentum, and weight decay},}\ }\href@noop {} {\  (\bibinfo {year}
  {2018})},\ \Eprint {http://arxiv.org/abs/1803.09820} {arXiv:1803.09820}
  \BibitemShut {NoStop}%
\bibitem [{\citenamefont {Kasieczka}\ \emph
  {et~al.}(2019{\natexlab{b}})\citenamefont {Kasieczka}, \citenamefont {Plehn},
  \citenamefont {Thompson},\ and\ \citenamefont
  {Russel}}]{kasieczka_gregor_2019_2603256}%
  \BibitemOpen
  \bibfield  {author} {\bibinfo {author} {\bibfnamefont {G.}~\bibnamefont
  {Kasieczka}}, \bibinfo {author} {\bibfnamefont {T.}~\bibnamefont {Plehn}},
  \bibinfo {author} {\bibfnamefont {J.}~\bibnamefont {Thompson}}, \ and\
  \bibinfo {author} {\bibfnamefont {M.}~\bibnamefont {Russel}},\ }\href
  {\doibase 10.5281/zenodo.2603256} {\enquote {\bibinfo {title} {Top quark
  tagging reference dataset},}\ } (\bibinfo {year}
  {2019}{\natexlab{b}})\BibitemShut {NoStop}%
\bibitem [{\citenamefont {Sjöstrand}\ \emph {et~al.}(2015)\citenamefont
  {Sjöstrand}, \citenamefont {Ask}, \citenamefont {Christiansen},
  \citenamefont {Corke}, \citenamefont {Desai}, \citenamefont {Ilten},
  \citenamefont {Mrenna}, \citenamefont {Prestel}, \citenamefont {Rasmussen},\
  and\ \citenamefont {Skands}}]{Sjostrand:2014zea}%
  \BibitemOpen
  \bibfield  {author} {\bibinfo {author} {\bibfnamefont {T.}~\bibnamefont
  {Sjöstrand}}, \bibinfo {author} {\bibfnamefont {S.}~\bibnamefont {Ask}},
  \bibinfo {author} {\bibfnamefont {J.~R.}\ \bibnamefont {Christiansen}},
  \bibinfo {author} {\bibfnamefont {R.}~\bibnamefont {Corke}}, \bibinfo
  {author} {\bibfnamefont {N.}~\bibnamefont {Desai}}, \bibinfo {author}
  {\bibfnamefont {P.}~\bibnamefont {Ilten}}, \bibinfo {author} {\bibfnamefont
  {S.}~\bibnamefont {Mrenna}}, \bibinfo {author} {\bibfnamefont
  {S.}~\bibnamefont {Prestel}}, \bibinfo {author} {\bibfnamefont {C.~O.}\
  \bibnamefont {Rasmussen}}, \ and\ \bibinfo {author} {\bibfnamefont {P.~Z.}\
  \bibnamefont {Skands}},\ }\bibfield  {title} {\enquote {\bibinfo {title} {{An
  Introduction to PYTHIA 8.2}},}\ }\href {\doibase 10.1016/j.cpc.2015.01.024}
  {\bibfield  {journal} {\bibinfo  {journal} {Comput. Phys. Commun.}\ }\textbf
  {\bibinfo {volume} {191}},\ \bibinfo {pages} {159--177} (\bibinfo {year}
  {2015})},\ \Eprint {http://arxiv.org/abs/1410.3012} {arXiv:1410.3012
  [hep-ph]} \BibitemShut {NoStop}%
%%CITATION = ARXIV:1410.3012;%%
\bibitem [{\citenamefont {de~Favereau}\ \emph {et~al.}(2014)\citenamefont
  {de~Favereau}, \citenamefont {Delaere}, \citenamefont {Demin}, \citenamefont
  {Giammanco}, \citenamefont {Lemaître}, \citenamefont {Mertens},\ and\
  \citenamefont {Selvaggi}}]{deFavereau:2013fsa}%
  \BibitemOpen
  \bibfield  {author} {\bibinfo {author} {\bibfnamefont {J.}~\bibnamefont
  {de~Favereau}}, \bibinfo {author} {\bibfnamefont {C.}~\bibnamefont
  {Delaere}}, \bibinfo {author} {\bibfnamefont {P.}~\bibnamefont {Demin}},
  \bibinfo {author} {\bibfnamefont {A.}~\bibnamefont {Giammanco}}, \bibinfo
  {author} {\bibfnamefont {V.}~\bibnamefont {Lemaître}}, \bibinfo {author}
  {\bibfnamefont {A.}~\bibnamefont {Mertens}}, \ and\ \bibinfo {author}
  {\bibfnamefont {M.}~\bibnamefont {Selvaggi}} (\bibinfo {collaboration}
  {DELPHES 3}),\ }\bibfield  {title} {\enquote {\bibinfo {title} {{DELPHES 3, A
  modular framework for fast simulation of a generic collider experiment}},}\
  }\href {\doibase 10.1007/JHEP02(2014)057} {\bibfield  {journal} {\bibinfo
  {journal} {JHEP}\ }\textbf {\bibinfo {volume} {02}},\ \bibinfo {pages} {057}
  (\bibinfo {year} {2014})},\ \Eprint {http://arxiv.org/abs/1307.6346}
  {arXiv:1307.6346 [hep-ex]} \BibitemShut {NoStop}%
%%CITATION = ARXIV:1307.6346;%%
\bibitem [{\citenamefont {Cacciari}\ \emph {et~al.}(2008)\citenamefont
  {Cacciari}, \citenamefont {Salam},\ and\ \citenamefont
  {Soyez}}]{Cacciari:2008gp}%
  \BibitemOpen
  \bibfield  {author} {\bibinfo {author} {\bibfnamefont {M.}~\bibnamefont
  {Cacciari}}, \bibinfo {author} {\bibfnamefont {G.~P.}\ \bibnamefont {Salam}},
  \ and\ \bibinfo {author} {\bibfnamefont {G.}~\bibnamefont {Soyez}},\
  }\bibfield  {title} {\enquote {\bibinfo {title} {{The anti-$k_t$ jet
  clustering algorithm}},}\ }\href {\doibase 10.1088/1126-6708/2008/04/063}
  {\bibfield  {journal} {\bibinfo  {journal} {JHEP}\ }\textbf {\bibinfo
  {volume} {04}},\ \bibinfo {pages} {063} (\bibinfo {year} {2008})},\ \Eprint
  {http://arxiv.org/abs/0802.1189} {arXiv:0802.1189 [hep-ph]} \BibitemShut
  {NoStop}%
%%CITATION = ARXIV:0802.1189;%%
\bibitem [{Note3()}]{Note3}%
  \BibitemOpen
  \bibinfo {note} {A comprehensive comparison between a wide range of machine
  learning approaches on this top tagging dataset is presented in Ref.~\cite
  {Kasieczka:2019dbj}, where an earlier version of ParticleNet is also
  included.}\BibitemShut {Stop}%
\bibitem [{\citenamefont {Xie}\ \emph {et~al.}(2017)\citenamefont {Xie},
  \citenamefont {Girshick}, \citenamefont {Doll{\'a}r}, \citenamefont {Tu},\
  and\ \citenamefont {He}}]{xie2017aggregated}%
  \BibitemOpen
  \bibfield  {author} {\bibinfo {author} {\bibfnamefont {S.}~\bibnamefont
  {Xie}}, \bibinfo {author} {\bibfnamefont {R.}~\bibnamefont {Girshick}},
  \bibinfo {author} {\bibfnamefont {P.}~\bibnamefont {Doll{\'a}r}}, \bibinfo
  {author} {\bibfnamefont {Z.}~\bibnamefont {Tu}}, \ and\ \bibinfo {author}
  {\bibfnamefont {K.}~\bibnamefont {He}},\ }\bibfield  {title} {\enquote
  {\bibinfo {title} {Aggregated residual transformations for deep neural
  networks},}\ }in\ \href {\doibase 10.1109/CVPR.2017.634} {\emph {\bibinfo
  {booktitle} {2017 IEEE Conference on Computer Vision and Pattern Recognition
  (CVPR)}}}\ (\bibinfo  {publisher} {IEEE},\ \bibinfo {address} {Honolulu, HI,
  USA},\ \bibinfo {year} {2017})\ pp.\ \bibinfo {pages}
  {5987--5995}\BibitemShut {NoStop}%
\bibitem [{\citenamefont {Butter}\ \emph {et~al.}(2019)\citenamefont {Butter}
  \emph {et~al.}}]{Kasieczka:2019dbj}%
  \BibitemOpen
  \bibfield  {author} {\bibinfo {author} {\bibfnamefont {A.}~\bibnamefont
  {Butter}} \emph {et~al.},\ }\bibfield  {title} {\enquote {\bibinfo {title}
  {{The Machine Learning Landscape of Top Taggers}},}\ }\href {\doibase
  10.21468/SciPostPhys.7.1.014} {\bibfield  {journal} {\bibinfo  {journal}
  {SciPost Phys.}\ }\textbf {\bibinfo {volume} {7}},\ \bibinfo {pages} {014}
  (\bibinfo {year} {2019})},\ \Eprint {http://arxiv.org/abs/1902.09914}
  {arXiv:1902.09914 [hep-ph]} \BibitemShut {NoStop}%
%%CITATION = ARXIV:1902.09914;%%
\bibitem [{\citenamefont {He}\ \emph {et~al.}(2016)\citenamefont {He},
  \citenamefont {Zhang}, \citenamefont {Ren},\ and\ \citenamefont
  {Sun}}]{DBLP:journals/corr/HeZR016}%
  \BibitemOpen
  \bibfield  {author} {\bibinfo {author} {\bibfnamefont {K.}~\bibnamefont
  {He}}, \bibinfo {author} {\bibfnamefont {X.}~\bibnamefont {Zhang}}, \bibinfo
  {author} {\bibfnamefont {S.}~\bibnamefont {Ren}}, \ and\ \bibinfo {author}
  {\bibfnamefont {J.}~\bibnamefont {Sun}},\ }\bibfield  {title} {\enquote
  {\bibinfo {title} {Identity mappings in deep residual networks},}\ }in\ \href
  {\doibase 10.1007/978-3-319-46493-0_38} {\emph {\bibinfo {booktitle}
  {Computer Vision -- ECCV 2016}}}\ (\bibinfo  {publisher} {Springer},\
  \bibinfo {address} {Cham},\ \bibinfo {year} {2016})\ pp.\ \bibinfo {pages}
  {630--645}\BibitemShut {NoStop}%
\end{thebibliography}%

\appendix

\section{Implementation details of ResNeXt-50}
\label{app:arch-resnext50}

The ResNeXt-50 model uses jet images as inputs. Each image is constructed from the constituent particles by projecting them onto a 2D grid of $64\times64$ pixels in size, corresponding to a granularity of 0.025 rad in the pseudorapidity-azimuth space. The intensity of each pixel is the sum of $p_{T}$ of all the particles within the pixel rescaled by the inverse of the jet $p_{T}$. 

The original 50-layer ResNeXt architecture \cite{xie2017aggregated} was developed for images of size $224\times224$ and a classification task with 1000 classes. To adapt to the smaller size of the jet images and the significantly fewer number of output classes, the number of channels in all but the first convolutional layers is reduced by a factor of 4, and a dropout layer with a drop probability of 0.5 is added after the global pooling layer. 

The network is implemented with Apache MXNet and trained with the Adam optimizer with a minibatch size of 256. The network is trained for 30 epochs, with a starting learning rate of 0.01, and subsequently reduced by a factor of 10 at the 10th and 20th epochs. A snapshot of the model is saved at the end of each epoch, and the model snapshot showing the best accuracy on the validation dataset is selected for the final evaluation. 

\section{Implementation details of P-CNN}
\label{app:arch-p-cnn}

The particle-level convolutional neural network (P-CNN) \cite{CMS-DP-2017-049} is a deep 1D CNN architecture customized for boosted jet tagging. Each input jet is represented as a sequence of particles with a fixed length of 100. The particles are organized in descending order of $p_T$. The sequence is padded with zeros if a jet has less than 100 particles and truncated if it has more than 100 particles. 

The P-CNN architecture is similar to the ResNet model \cite{he2016deep,DBLP:journals/corr/HeZR016} for image classification but uses 1D convolution instead. It features a total of 14 convolutional layers, all with a kernel size of 3. The number of channels for the 1D convolutions is either 32, 64, or 128. The convolutions are followed by a global pooling, then by a fully connected layer of 512 units with ReLU activation and a dropout layer with a drop rate of 0.5, before producing the classification output. 

The network is implemented with Apache MXNet and trained with the Adam optimizer with a minibatch size of 1024. The network is trained for 30 epochs, with a starting learning rate of 0.001, and subsequently reduced by a factor of 10 at the 10th and 20th epochs. A snapshot of the model is saved at the end of each epoch, and the model snapshot showing the best accuracy on the validation dataset is selected for the final evaluation.

\end{document}